%%%%%%%%%%%%%%%%%

\input harvmac
\noblackbox
%%% Figures
\newcount\figno
\figno=0
\def\fig#1#2#3{
\par\begingroup\parindent=0pt\leftskip=1cm\rightskip=1cm\parindent=0pt
\baselineskip=11pt
\global\advance\figno by 1
\midinsert
\epsfxsize=#3
\centerline{\epsfbox{#2}}
\vskip 12pt
\centerline{{\bf Figure \the\figno:} #1}\par
\endinsert\endgroup\par}
\def\figlabel#1{\xdef#1{\the\figno}}

\def\AdSS5{$AdS_5$}
\def\AdS5s5{$AdS_5 \times S^5$}

\def\NSNS{{$NS\otimes NS$}}
\def\RR{{$R\otimes R$}}
\def\ZZ {{\bf Z}}

\def\calR{{\cal R}}

\def\det{\hbox{\rm det}}

\def\G(#1){\Gamma(#1)}

\def\C|#1{{\cal #1}     }
\def\(#1#2){(\zeta_#1\cdot\zeta_#2)}

\def\np#1#2#3{Nucl. Phys. {\bf B#1} (#2) #3}
\def\pl#1#2#3{Phys. Lett. {\bf B#1} (#2) #3}

\def\prd#1#2#3{Phys. Rev. {\bf D#1} (#2) #3}

%%% Paragraphs

%%% special math symbols
\font\cmss=cmss10
\font\cmsss=cmss10 at 7pt
\def\rlx{\relax\leavevmode}
\def\inbar{\vrule height1.5ex width.4pt depth0pt}
\def\IC{\relax\,\hbox{$\inbar\kern-.3em{\rm C}$}}
\def\IN{\relax{\rm I\kern-.18em N}}
\def\IP{\relax{\rm I\kern-.18em P}}
\def\ZZ{\rlx\leavevmode\ifmmode\mathchoice{\hbox{\cmss Z\kern-.4em Z}}
 {\hbox{\cmss Z\kern-.4em Z}}{\lower.9pt\hbox{\cmsss Z\kern-.36em Z}}
 {\lower1.2pt\hbox{\cmsss Z\kern-.36em Z}}\else{\cmss Z\kern-.4em
 Z}\fi}
%%% misc.
\def\IZ{\relax\ifmmode\mathchoice
{\hbox{\cmss Z\kern-.4em Z}}{\hbox{\cmss Z\kern-.4em Z}}
{\lower.9pt\hbox{\cmsss Z\kern-.4em Z}}
{\lower1.2pt\hbox{\cmsss Z\kern-.4em Z}}\else{\cmss Z\kern-.4em
Z}\fi}

\def\narrowplus{\kern -.04truein + \kern -.03truein}
\def\narrowminus{- \kern -.04truein}
\def\narrowminussub{\kern -.02truein - \kern -.01truein}
\font\ninerm=cmr9

\def\kh{K\"{a}hler}
\def\half{{1\over 2}}

\def\e{{\epsilon}}

\def\frac#1#2{{#1\over #2}}

\def\com#1#2{{ \left[ #1, #2 \right] }}

\def\IZ{\relax\ifmmode\mathchoice
{\hbox{\cmss Z\kern-.4em Z}}{\hbox{\cmss Z\kern-.4em Z}}
{\lower.9pt\hbox{\cmsss Z\kern-.4em Z}}
{\lower1.2pt\hbox{\cmsss Z\kern-.4em Z}}\else{\cmss Z\kern-.4em
Z}\fi}
\def\IB{\relax{\rm I\kern-.18em B}}
\def\IC{{\relax\hbox{$\inbar\kern-.3em{\rm C}$}}}
\def\ID{\relax{\rm I\kern-.18em D}}
\def\IE{\relax{\rm I\kern-.18em E}}
\def\IF{\relax{\rm I\kern-.18em F}}
\def\IG{\relax\hbox{$\inbar\kern-.3em{\rm G}$}}
\def\IGa{\relax\hbox{${\rm I}\kern-.18em\Gamma$}}
\def\IH{\relax{\rm I\kern-.18em H}}
\def\II{\relax{\rm I\kern-.18em I}}
\def\IK{\relax{\rm I\kern-.18em K}}
\def\IP{\relax{\rm I\kern-.18em P}}
%\def\IX{\relax{\rm X\kern-.01em X}}
%this doesn't work

\font\cmss=cmss10 \font\cmsss=cmss10 at 7pt
\def\IR{\relax{\rm I\kern-.18em R}}

\def\f{\psi}

\def\1{{\bf 1}}
\def\3{{\bf 3}}
\def\7{{\bf 7}}
\def\2{{\bf 2}}
\def\8{{\bf 8}}

\def\com#1#2{{ \left[ #1, #2 \right] }}

%

%
%       \eqn\label{a+b=c}       gives displayed equation, numbered
%                               consecutively within sections.
%     \eqnn and \eqna define labels in advance (of eqalign?)
%
\def\eqnn#1{\xdef #1{(\secsym\the\meqno)}\writedef{#1\leftbracket#1}%
\global\advance\meqno by1\wrlabeL#1}
\def\eqna#1{\xdef #1##1{\hbox{$(\secsym\the\meqno##1)$}}
\writedef{#1\numbersign1\leftbracket#1{\numbersign1}}%
\global\advance\meqno by1\wrlabeL{#1$\{\}$}}
\def\eqn#1#2{\xdef #1{(\secsym\the\meqno)}\writedef{#1\leftbracket#1}%
\global\advance\meqno by1$$#2\eqno#1\eqlabeL#1$$}

\def\Im{\rm Im}

\def\tr{{\rm tr}}

\def\half{{\textstyle {1 \over 2}}}

\def\lam16{\lambda^{16}}

\def\lr { \lref}

%\def\xxx#1 {{hep-th/#1}}
%\def\npb#1(#2)#3 { Nucl. Phys. {\bf B#1} (#2) #3 }
%\def\rep#1(#2)#3 { Phys. Rept.{\bf #1} (#2) #3 }
%\def\plb#1(#2)#3{Phys. Lett. {\bf #1B} (#2) #3}
%\def\prl#1(#2)#3{Phys. Rev. Lett.{\bf #1} (#2) #3}
%\def\physrev#1(#2)#3{Phys. Rev. {\bf D#1} (#2) #3}
%\def\ap#1(#2)#3{Ann. Phys. {\bf #1} (#2) #3}
%\def\rmp#1(#2)#3{Rev. Mod. Phys. {\bf #1} (#2) #3}
%\def\cmp#1(#2)#3{Comm. Math. Phys. {\bf #1} (#2) #3}
%\def\mpl#1(#2)#3{Mod. Phys. Lett. {\bf #1} (#2) #3}
%\def\ijmp#1(#2)#3{Int. J. Mod. Phys. {\bf A#1} (#2) #3}

% References for susy constraints

\lr\rvb{N. Berkovits and C. Vafa, {\it N=4 Topological Strings}, 
hep-th/9407190, \np{433}{1995}{123}.}
\lr\rvo{H. Ooguri and C. Vafa, {\it All Loop N=2 String Amplitudes}, 
hep-th/9505183, \np{451}{1995}{121}.}
\lr\rstrings{N. Berkovits, {\it Generalization of the $R^4$ conjecture},
 proceedings of  Strings '98\semi 
http://www.itp.ucsb.edu/online/strings98/berkovits/.}
\lr\rWSD{E. Witten, {\it String Theory Dynamics in Various Dimensions},
hep-th/9503124, \np{443}{1995}{85}.}
\lr\rPKT{P. K. Townsend, {\it The Elven-dimensional Supermembrane Revisited},
hep-th/9501068, \pl{350}{1995}{184}.}
\lr\rkirit{ B. Pioline and E. Kiritsis, {\it  On $R^4$ Threshold Corrections in
IIB
String Theory and $(P, Q)$ String Instantons}, hep-th/9707018,
\np{508}{1997}{509}.}
\lr\rpioline{ B. Pioline, {\it A Note On Nonperturbative $R^4$ Couplings},
hep-th/9804023.}
\lr\rberk{N. Berkovits, {\it Construction of $R^4$ Terms in N=2 D=8
Superspace},
hep-th/9709116, \np{514}{1998}{191}. }
\lr\russo{J. G. Russo, {\it Construction of SL(2,Z) Invariant Amplitudes in
Type
IIB Superstring Theory}, hep-th/9802090; {\it An Ansatz For A Nonperturbative
Four
Graviton Amplitude In Type IIB Superstring Theory}, hep-th/9707241,
\pl{417}{1998}{253}.}
\lr\kehagias{A. Kehagias and H. Partouche, {\it The Exact Quartic Effective
Action
 For The Type IIB Superstring}, hep-th/9710023, \pl{422}{1998}{109}.  }
\lr\terras{ A. Terras, {\it Harmonic Analysis on Symmetric Spaces and
Applications I}, Springer-Verlag (New York) 1985. }
\lr\berkovafa{N. Berkovits and C. Vafa, {\it Type IIB $R^4 H^{4g-4}$
Conjectures}, hep-th/9803145. }
\lr\rmIIB{P. Aspinwall, {\it Some Relationships Between Dualities in
String Theory}, hep-th/9508154, Nucl. Phys. Proc. Suppl. {\bf 46}
(1996) 30\semi J. Schwarz, {\it The Power of M Theory},
hep-th/9510086, \pl{367}{1996}{97}. }
\lr\cjs {E. Cremmer, B. Julia and J. Scherk, {\it Supergravity
Theory in
Eleven Dimensions}, Phys. Lett. {\bf 76B} (1978) 409.}
\lr\minasduff{M.J.~Duff, J.T.~Liu and R.~Minasian, {\it
Eleven-dimensional Origin
of String-String Duality: A One Loop Test}, hep-th/9506126,
Nucl.  Phys. {\bf  452B} (1995) 261.}
\lr\vafawitt{C.~Vafa and E.~Witten, {\it A One Loop Test of
String Duality},
  hep-th/9505053, Nucl.  Phys. {\bf B447} (1995) 261.}
\lr\sethia{S.  Sethi,  S.  Paban and M.  Stern, {\it Constraints
{}From Extended Supersymmetry in Quantum Mechanics}, hep-th/9805018;
{\it
Supersymmetry and Higher Derivative Terms in the Effective
Action
of Yang--Mills Theories},  hep-th/9806028.}
\lr\stromingerb{A. Strominger, {\it Loop Corrections to the Universal
Hypermultiplet},
hep-th/9706195.}\lr\ferrarab{I. Antoniadis, S. Ferrara, R. Minasian and K.S.
Narain,
{\it $R^4$ couplings in M and type II theories},
hep-th/9707013.}
\lr\knn{J.Koplik, A.Neveu, S.Nussinov,  {\it Some aspects of the planar
perturbation series}, Nucl. Phys.  {\bf B123} (1977)  109.}
\lr\ew{E.Witten,  {\it Current algebra theorems for the U(1) \lq Goldstone
boson'}, Nucl. Phys.  {\bf B156}  (1979)  269;  {\it Instantons, the quark
model
and the $1/N$ expansion}, Nucl. Phys. {\bf B149}
(1979) 285.}
\lr\jr{R.Jackiw, C.Rebbi, Phys. Rev. {\bf D14}, (1976), 517.}
\lr\hh{G.Horowitz, H.Ooguri, hep-th/9802116 .}
\lr\kslvz{S.Kachru, E.Silverstein, {\it 4-D conformal Theories and Strings on
Orbifolds}, hep-th/9802183;
A.Lawrence, N.Nekrasov, C.Vafa, {\it On Conformal Field Theories in Four
Dimensions}, hep-th/9803015. }
\lr\thorn{C.B.  Thorn, in Sakharov Conference on Physics, Moscow, (91),447.}
\lr\dpsgketc{S.Gubser, I.Klebanov, A.Peet, Phys. Rev. {\bf D54}, (1996), 3915,
hep-th/9602135 ;
 A.A.Tseytlin, I.Klebanov, Nucl. Phys. {\bf B475}, (1996), 179, hep-th/9604166
;
 S.Gubser, I.Klebanov, Phys. Lett. {\bf B413}, (1997), 41, hep-th/9708005; and
references cited in \juan.}
\lr\juan{J.  Maldacena, {\it   The large $N$ limit of
superconformal field
theories and supergravity}, hep-th/971120.}
\lr\bvz{M.Bershadsky, Z.Kakushadze, C.Vafa, {\it String Expansion as Large N
Expansion of Gauge Theories}, hep-th/9803076;
M.Bershadsky, A.Johansen, {\it Large N Limit of Orbifold Field Theories},
hep-th/9803249.}
\lr\gk{S.S. Gubser and I.R.  Klebanov, {\it Absorption by branes
and Schwinger
terms in the world volume theory}, hep-th/9708005.}
\lr\gkp{S.S. Gubser,  I.R.  Klebanov and A.M.  Polyakov, {\it
Gauge theory
correlators from non-critical string theory},  hep-th/9802109.}
\lr\wittone{E.  Witten, {\it Anti de Sitter Space and Holography},
hep-th/9802150.}
\lr\tHS{G. 't Hooft, {\it Dimensional Reduction in Quantum Gravity},
gr-qc/9310006; L.Susskind, {\it The World as a Hologram},
J. Math. Phys. 36, 6377 (1995), hep-th/9409089.}
\lr\greenvan{M.B.  Green and P. Vanhove,
{\it D-instantons, Strings and M-theory},
Phys. Lett. {\bf B408} (1997) 122, hep-th/9704145.}
\lr\greengutvan{M.B.~Green,  M.~Gutperle and P.~Vanhove, {\it One Loop in
Eleven-Dimensions}, hep-th/9706175, \pl{409}{1997}{177}.}
\lr\greengutkwon{M.B.~Green,  M.~Gutperle and H.~Kwon, {\it Sixteen Fermion
 and Related
Terms in M theory on $T^2$}, hep-th/9710151.}
\lr\wittfive{E. Witten, {\it Five-brane Effective Action in M Theory}, 
 hep-th/9610234,
 J. Geom. Phys. {\bf 22} (1997) 103.}
\lr\nilsson{B.E.W. Nilsson and A.  Tollsten, {Supersymmetrization of
$\zeta(3) R_{\mu\nu\tau\sigma}^4$ in superstring theories}, Phys. Lett
{\bf 181B} (1986) 63.}
\lr\greengut{M.B.~Green and M.~Gutperle, {\it Effects of D-instantons},
hep-th/9701093, \np{498}{1997}{195}.}
\lr\greenschwarz{M.B.  Green and J.H. Schwarz, {\it Supersymmetric
Dual String Theory (II).  Loops and Renormalization}, Nucl.  Phys.
{\bf B198} (1982) 441.}
\lr\osborn{  J.  Erdmenger and H.  Osborn, {\it
Conformally covariant differential operators: Symmetric tensor
fields}, gr-qc/9708040; Class. Quantum Grav. {\bf 15} (1998) 273.}
\lr\eguchi{T.  Eguchi, {\it S Duality and Strong Coupling Behavior of Large N
Gauge Theories with N=4 Supersymmetry}, hep-th/9804037.}
\lr\schwarzwest{J.H.  Schwarz and P.C.  West, {\it  Symmetries and
Transformations of Chiral $N=2$, $D=10$ Supergravity}, Phys. Lett.
 {\bf 126B} (1983) 301.}
\lr\greenschwarza{M.B.  Green and J.H.  Schwarz, {\it Extended
 Supergravity in Ten Dimensions}, Phys.  Lett.  {\bf 122B} (1983) 143.}
\lr\schwarza{J.H.  Schwarz, {\it Covariant Field Equations of Chiral
$N=2$, $D=10$ Supergravity},  Nucl. Phys.  {\bf B226} (1993) 269.}
\lr\howest{P.S.  Howe and P.C. West, {\it The Complete $N=2$ $D=10$
Supergravity}, Nucl.  Phys.  {\bf B238} (1984) 181.}
\lr\matrixth{T.~Banks, W.~Fischler, S.H.~Shenker and  L.~Susskind,
{\it M
Theory As A Matrix Model: A Conjecture}, hep-th/9610043,
\prd{55}{1997}{5112}.}
\lr\gw{D.J.~Gross and E.~Witten, {\it Superstring modifications of
Einstein's
    equations}, \npb277(1986)1.}
\lr\gris{M.T.~Grisaru , A.E.M~Van de Ven and D.~Zanon, {\it
Two-dimensional
supersymmetric sigma models on Ricci flat Kahler manifolds are not
finite},
\npb277(1986)388 ; {\it Four loop divergences for the N=1 supersymmetric
nonlinear sigma model in two-dimensions}, \npb277(1986)409.}
\lr\jackreb{R.  Jackiw and C.  Rebbi, {\it Spinor analysis of
Yang--Mills theory}, Phys.  Rev.  {\bf D16} (1977) 1052.}
\lr\greengutc{M.B.~Green and M.~Gutperle, {\it D-particle Bound States and the
D-instanton Measure}, hep-th/9711107, {\bf JHEP01} (1998) 005.}
\lr\f{D.Z.Freedman, S.D.Mathur, A.Matusis, L.Rastelli, {\it Correlation
Functions in the CFT(D)/ADS(D+1) Correspondence}, hep-th/9804058 .}
\lr\greencargese{M.B.  Green, {\it Connections between M theory and 
superstrings}, Proceedings of the 1997 Advanced Study Institute on Strings, 
Branes and Dualities, Cargese, \hep-th/9712195; Nucl. Phys. Proc. Suppl. 
{\bf 68} (1998) 242.}

\Title{\vbox{\baselineskip12pt
\hbox{hep-th/9808061}
\hbox{DAMTP-98-96, IASSNS--HEP--98/70}
\hbox{}
}}
{\vbox{
\centerline{Supersymmetry Constraints on Type IIB Supergravity}
  }}
\centerline{Michael B.  Green\foot{ M.B.Green@damtp.cam.ac.uk}}
\vskip 0.05in
\centerline{\it Department of Applied Mathematics and
Theoretical Physics}
\centerline{\it Cambridge University}\centerline{\it Cambridge CB3 9EW, UK}
\smallskip
\centerline{ and}
\smallskip
\centerline{ Savdeep Sethi\foot{ sethi@sns.ias.edu}}
\vskip 0.05in
\medskip\centerline{\it School of Natural Sciences}
\centerline{\it Institute for Advanced Study}\centerline{\it
Princeton, NJ
08540, USA}

\vskip 0.3in

Supersymmetry is used to derive conditions on higher derivative terms
in the effective action of type IIB supergravity. 
Using these conditions, we are
able to
prove earlier conjectures that certain modular invariant interactions of order  
$(\alpha')^3$
relative to the Einstein--Hilbert term are proportional to eigenfunctions of
the
Laplace operator on the fundamental domain of $SL(2,\IZ)$. 
We also discuss how
these arguments generalize to terms of higher order in $\alpha'$, as well as to
compactifications of supergravity.

%\draftmode
\vskip 0.1in
\Date{8/98}

\newsec{Introduction}

Despite the intense interest in the structure of M theory, the general
constraints imposed by  supersymmetry
have not been systematically investigated. At low energies in eleven 
dimensions, M theory should be well approximated by eleven-dimensional
supergravity \refs{\cjs}. However, eleven-dimensional supergravity is not 
a consistent quantum theory and new ingredients are needed which modify the 
ultraviolet properties of the theory.   While eventually we hope to have a 
 microscopic formulation 
of M theory, it is  
interesting to unravel the
extent to which its structure is constrained simply 
by general symmetry
principles. For example, the 
cancellation of chiral gauge and gravitational anomalies induced on
the  five-brane
leads immediately to a term in the effective action of the form,
\eqn\firsd{ \int C^{(3)} \wedge X_8(R),}
where $X_8$ is an eight-form constructed from curvatures, and $C^{(3)}$ is
the three form tensor field \refs{\vafawitt,\minasduff}. This term is eighth
order in derivatives compared to
classical terms in the effective action which are second order. As usual, the
order
in a momentum expansion counts
the number of derivatives plus twice the number of
fermions. Clearly, we can generate many more terms needed for a supersymmetric
effective action by acting with the lowest order supersymmetry transformations
on these higher derivative terms. Some of these terms have been deduced from
duality arguments.  Furthermore, as soon as there are 
eight derivative terms in the
effective action, there will be sixth order modifications to the classical
supersymmetry transformations. The action is then no longer invariant under
supersymmetry unless we add yet higher order terms to the effective action.

Ideally, it would be possible to describe the theory in a manner
that is independent of the background. The moduli
of an arbitrary compactification would then emerge from components of the
tensor
bosonic fields.  However, in practice it is only feasible to define the
effective action with respect to a given moduli space.  The simplest
example with substantial structure is the
 Poincar\'e invariant ten-dimensional  background appropriate to the type
IIB superstring,  which has moduli space $SL(2,\IR)/U(1)$. The type IIB string
arises by compactifying M theory on a $T^2$, where the volume of the
torus is taken to zero \rmIIB. The complex structure of the torus becomes the
complex coupling $\tau = C^{(0)} + i e^{-\phi}$ of the IIB theory, where
 $C^{(0)}$ is the Ramond--Ramond ($R\otimes R$) scalar and $\phi$ is the
dilaton.

The type IIB  effective action
is expressed as an expansion in powers of $\alpha'$ with the classical theory
defined by an `action' $S^{(0)}$ of order $(\alpha')^{-4}$ 
\refs{\greenschwarza,\schwarzwest,
\schwarza,\howest}.   It is well-known that the self-duality constraint
on the five-form field strength in the type IIB theory
presents
an obstacle to actually writing a globally defined covariant action 
$S^{(0)}.$\foot{
This issue has been  
thoroughly discussed in \wittfive.} However, the analysis
in our paper will actually  only involve the field
equations. We will use terminology appropriate for a theory with an action, but
merely as a shorthand method
of packaging these equations. We could avoid this problem by compactifying the
type
IIB theory on a circle. The classical moduli space would then be 
$SL(2,\IR)/U(1)\times \IR$.

The supersymmetry transformations on an arbitrary field $\Psi$ will be
expressed as the series,
\eqn\susyexp{\delta_\epsilon \Psi
=\left( \delta^{(0)} + \alpha' \delta^{(1)}
+\ldots + (\alpha')^n \delta^{(n)} + \ldots \right)\Psi,}
while the effective action has the following expansion,
\eqn\effacs{S = S^{(0)} + \alpha'  S^{(1)} + \ldots + (\alpha')^n S^{(n)} +
\ldots.}
A factor of $(\alpha')^{-4}$ has been absorbed into the definition of
$S$ and $S^{(n)}$.

 In principle, the action can be constructed
by a Noether procedure which imposes the conditions,
\eqn\noethn{ \left( \sum_{m=0}^r (\alpha')^m \delta^{(m)} \right) \sum_{n=o}^r
(\alpha')^n S^{(n)}  = 0 ,}
order by order in $\alpha'$. There are no $n=1$ or $n=2$ terms at tree-level
or one-loop, and these terms are not expected to appear at all
in \effacs. Therefore,
the first corrections are the terms of order
$(\alpha')^{3}$ relative to $S^{(0)}$. These terms are eighth order in
derivatives.

  In practice, building the complete effective action from scratch using this
Noether method is extremely complicated even for a low number of derivatives.
However, we will show in this paper that  the exact form of special 
classes
of M theory or type IIB interactions can be uniquely 
determined in this manner.  
A similar analysis has recently been used to
obtain powerful constraints on
maximally supersymmetric Yang--Mills theories \sethia. The lesson
to be drawn from that analysis is that the constraints imposed by supersymmetry
are most easily exhibited by studying the variation of terms in the effective
action
with the maximal number of fermionic fields.

Among other issues, one of our aims will be to establish the validity of  some
conjectured higher derivative interactions in the effective action.  
An example of such a term is the interaction,
  $\int \sqrt g f^{(0,0)}(\tau,\bar\tau) \calR^4$, where
$\calR^4$ is a particular contraction of four Weyl curvatures \greengut.
The $SL(2,\IZ)$ symmetry of the IIB theory requires that
$f^{(0,0)}(\tau,\bar \tau)$ be  a modular function of the complex
scalar field $\tau$ and its complex conjugate, $\bar \tau$.
It was noted in \greengut\ that $f^{(0,0)}$  is an
eigenfunction of the Laplace operator on the $SL(2,\IZ)$
moduli space with eigenvalue
$3/4$, 
\eqn\lapone{\nabla^2 f^{(0,0)} \equiv
4\tau_2^2 {\partial\over \partial\tau}{\partial\over
\partial\bar\tau} f^{(0,0)} = {3\over 4} f^{(0,0)}.}
This equation  has the  solution (see, for example, \terras),
\eqn\resfo{f^{(0,0)} = \sum_{(m,n)\ne (0,0)}{\tau_2^{3/2} \over |m+n \tau|^3},}
which is the unique solution, up to an arbitrary overall constant factor, 
for a choice of
asymptotic behavior near the boundary $\tau_2\to  \infty$
of the fundamental domain of 
$SL(2, \IZ)$. The  asymptotic behavior is determined by the weak coupling
expansion of $f^{(0,0)}$, where $\tau_2 = e^{-\phi}$ is large,  which possesses 
a
tree-level and one-loop term but
no other perturbative corrections.  In addition, there are an infinite
number of D-instanton corrections.\foot{Some supplementary evidence for the
expression \resfo, based
on linearized supersymmetry, is given in \refs{\rberk,\rpioline}.}    
Another term of the same order is the
sixteen dilatino term,
$f^{(12,-12)}\lambda^{16}$, where  the dilatino, $\lambda^a$ ($a=1,
\cdots, 16$),  is a complex $SO(9,1)$ Weyl spinor. This term  was discussed
in \greengutkwon\
where it was argued that,
\eqn\ftwel{f^{(12,-12)}(\tau,\bar\tau) =  D ^{12} f^{(0,0)}.}
The modular covariant derivative $D$ will be defined in
the next section.  This
 means that $f^{(12,-12)}$ should also be an
eigenfunction of the Laplace operator, but it now transforms with
the non-trivial holomorphic and anti-holomorphic modular weights indicated
by the superscripts.

More generally, there are many other
terms in $S^{(3)}$ that are related to the
$\calR^4$ term by supersymmetry at the linearized level \refs{\greengutkwon,
\kehagias}. The moduli dependence of these terms is packaged into a variety of
modular forms,  $f^{(w,-w)}(\tau,\bar\tau)$.
In section two,  we will  review how linearized
supersymmetry leads to the existence of all the terms in $S^{(3)}$ once
the presence of the $\calR^4$ term is assumed. However, linearized
supersymmetry
is certainly not powerful enough to determine the moduli dependent
coefficients,
$f^{(w,-w)}$.

In section three, we will use the full nonlinear supersymmetry   to
determine the nonholomorphic modular forms.  This
 requires a detailed analysis of the lowest order
supersymmetry transformations which generally mix
 all the terms in $S^{(3)}$. We will make  a judicious choice of
terms to consider in order to encounter minimal complications.  Not
surprisingly as in the cases of \sethia, it turns out that the terms with the
maximal number of fermions are the appropriate ones for this purpose.
The particular
terms we will consider are $f^{(12,-12)}\lambda^{16}$ and $
f^{(11,-11)} \lambda^{15} \psi^*_\mu$, where the latter is a piece of the
$\lambda^{14} \hat G$
term. Our notation and conventions are explained in Appendix A -- a hat on a
field strength indicates that it  includes
certain fermion bilinears in its definition in order to make it
`supercovariant.'

In addition, we will be forced to consider terms arising from
$O((\alpha')^3)$ supersymmetry transformations acting on the classical action.
These terms from $S^{(0)}$ mix under a supersymmetry variation with the
relevant
terms in $S^{(3)}$.  For our particular purpose, it will be
important to consider  a $\lambda^2\lambda^{*2}$ term in the IIB action
that has not to our knowledge been given explicitly in the
literature. The form  
of this term, including its precise normalization, 
is determined by supersymmetry 
in Appendix B.
By requiring invariance of the action at  order $(\alpha')^3$
together with closure of the supersymmetry algebra, we will be able
to determine  certain modifications to the supersymmetry transformations,
encoded in $\delta^{(3)}$, as well as the precise coefficients of the terms in
$S^{(3)}$ under investigation. As usual, the supersymmetry algebra
only closes with the use of the equations of motion.

In particular, we will find that the coefficients
$f^{(11,-11)}$ and $f^{(12,-12)}$ do indeed satisfy the appropriate
Laplace equations, proving the earlier conjectures about these modular forms.
Furthermore, once these functions have been determined the other terms in 
$S^{(3)}$   that 
are related to these by linearized supersymmetry, including the $\calR^4$
term,  follow without the need
for detailed analysis.

There
have
also been generalizations of the $\calR^4$ conjecture to an infinite series of
higher
order terms in the type IIB effective action \refs{\berkovafa, \russo}. In 
section 4 we outline how our technique can be extended to determine the
coefficients of some of these higher derivative interactions.  
We demonstrate how the constraints imposed by supersymmetry on these higher 
derivative interactions can be obtained but we do not carry through the
detailed calculation, which would be reasonably complicated.
It would  be very interesting to extend this analysis 
to compactified supergravity to prove and generalize, for example,  
conjectures like those in \rkirit.

\newsec{Higher Order Terms in the Type IIB Effective Action}

\subsec{Linearized supersymmetry and terms of order $(\alpha')^{3}$}

The existence of a large number of interactions in the IIB theory that are
related to the ${\cal R}^4$ interaction can be motivated
very simply by using
linearized supersymmetry.  This can be implemented by
packaging the physical fields or their field strengths into a constrained
superfield $\Phi(x,\theta)$ where $\theta^a$  ($a=1, \dots,16$)
is  a complex Grassmann coordinate that transforms as a Weyl spinor of
$SO(9,1)$.  This superfield satisfies the constraints,
\eqn\consphi{{\bar D}\Phi = 0,\qquad {\bar D}^4 {\bar \Phi} =0=  D^4 \Phi,}
where the first constraint is a chirality condition that ensures that $\Phi$ is
independent of $\theta^*$. The last two constraints imply that the components
of $\Phi$ satisfy the free field equations.  The superfield terminates after
the
$\theta^8$ term and has a component expansion that takes the form,\foot{We are 
using the usual convention that $\gamma^{\mu_1 \dots \mu_p}$ is the 
antisymmetrized product of $p$ gamma matrices, normalized so that 
$\gamma^{\mu_1\dots \mu_p} \equiv \gamma^{
\mu_1} \dots \gamma^{\mu_p}$ when $\mu_1 \ne \dots \ne \mu_p$.}
\eqn\expphi{\eqalign{\Phi =& \tau + i\bar \theta^* \lambda + \hat G_{\mu\nu\rho}
\bar\theta^* \gamma^{\mu\nu\rho} \theta + \cdots +  \calR_{\mu\sigma\nu\tau}
  \bar\theta^* \gamma^{\mu\nu\rho} \theta \bar\theta^*
\gamma^{\sigma\tau}_{\ \ \ \rho} \theta  \cr &
 +\partial_\mu 
\hat F_{5\, \nu\rho\sigma\tau\omega}\, \bar\theta^* \gamma^{\mu\nu\rho} 
\theta \bar\theta^*
\gamma^{\sigma\tau\omega} \theta +
 \cdots  + \theta^8 \partial^4 \bar
\tau.\cr}}
The  symbol  $\hat G_{\mu\nu\rho}$,
$\mu,\nu,\rho = 0, \ldots, 9$, denotes the `supercovariant' combination of
$G$ and fermion bilinears defined in Appendix A, where 
 $G_{\mu\nu\rho}$ and $G_{\mu\nu\rho}^*$ are complex combinations of the field
strengths
of the \RR\ and \NSNS\ (Neveu-Schwarz -- Neveu-Schwarz) two-form potentials.
The four-theta terms are  $\calR$,  the
Weyl curvature, 
and  $F_{5\, \rho_1\cdots\rho_5}$, which is the field strength of the
fourth-rank \RR\ potential.  The gamma matrices with world indices are defined 
by
$\gamma^\mu = e^\mu_m \gamma^m$, where $m=0,\cdots,9$ is the $SO(9,1)$
tangent-space index and $e^\mu_m$ is the inverse zehnbein.
A barred Weyl spinor, such as $\bar\theta$, is defined by
\eqn\bardef{\bar\theta_a \equiv \theta^*_b (\gamma^0)_{ba}.}
We have been sketchy about the precise coefficients
in \expphi\ since their
values   will not concern us.
The interactions that will be of interest in the next section are those that
arise by integrating a function of $\Phi$ over the sixteen
 components of $\theta$. In Einstein frame, this leads to  interaction terms
 of the following form,
\eqn\nonpert{\eqalign
{S^{(3)} = & (\alpha')^{3} \int d^{10} x d^{16}  \theta\, \det\, e \,F[\Phi]
+ {\rm c.c.} \cr  =&(\alpha')^{3}
\int d^{10}x\, \det\, e\, \left(
f^{(12,-12)} \lambda^{16} + f^{(11,-11)} \hat G \lambda^{14} + \ldots
\right .\cr
&  \left.
+ f^{(8,-8)} \hat G^8 +\ldots + f^{(0,0)} \calR^4 + \ldots +
f^{(-12,12)} \lambda^{* \, 16}\right),\cr}}
where  $\det\, e = \det\, e_\mu^m$ is the determinant
of the zehnbein.
The $SL(2,\IZ)$ symmetry of the IIB theory requires that
all the funtions, $f^{(w,-w)}(\tau,\bar\tau)$,
are modular forms with
holomorphic and anti-holomorphic weights as indicated in the
superscripts. Many terms have been hidden in the ellipsis in \nonpert.

We will mainly consider the first two terms  in parentheses  on the
right-hand-side of  \nonpert\ where we are using the precise notation,
\eqn\defnot{(\lambda^r)_{a_{r+1} \cdots a_{16}} \equiv {1\over r!}
\epsilon_{a_1\cdots a_{16}} \lambda^{a_1} \dots \lambda^{a_r}, }
so that,
\eqn\definv{\lambda^{16} =
{1\over 16!}\epsilon_{a_1 \dots a_{16}}
 \lambda^{a_1} \dots \lambda^{a_{16}},}
and
\eqn\ghatlam{\eqalign{\hat G \lambda^{14} & \equiv \hat  G_{\mu\nu\rho}
(\gamma^{\mu\nu\rho}\gamma^0)_{a_{15} a_{16}}  (\lambda^{14})_{a_{15} a_{16}},
\cr &=
{1\over 14!}\hat G_{\mu\nu\rho}(\gamma^{\mu\nu\rho}\gamma^0)_{a_{15}a_{16}}\,
\epsilon_{a_1 \dots a_{16}}   \lambda^{a_1} \dots
\lambda^{a_{14}}  .}}
Later we will make use of the simple identities,
\eqn\identb{ \eqalign{(\lambda^{14})_{ab}\, \lambda^c &=
(\lambda^{15})_b \, \delta_{a}^c -
(\lambda^{15})_a\, \delta_{b}^c, \cr
(\lambda^{15})_a\, \lambda^b &= \delta_{a}^b \,\lambda^{16}, \cr
 (\lambda)^{15}_a \, \lambda^a &= 16 \lambda^{16},}}
and
\eqn\identa{ (\lambda^{14})_{ab}\, \lambda_c \lambda_d =
\lambda^{16}\, (\delta_{ac}\delta_{bd} - \delta_{ad} \delta_{bc}).}

\subsec{Modular covariance}

The various coefficient functions  in the effective action are
 $(w,\hat w)$ forms, where $w$ refers to the holomorphic modular weight and
 $\hat w$ to the anti-holomorphic  modular weight.  A nonholomorphic
modular form $F^{(w,\hat w)}$ transforms as,
\eqn\ftranss{F^{(w,\hat w)} \to F^{(w,\hat w)}\, (c\tau + d)^w
(c\bar \tau + d)^{\hat w},}
under the $SL(2,\IZ)$ transformation taking,
\eqn\tautrans{\tau \to {a\tau + b \over c\tau+d}.}
Equation \ftranss\  describes a $U(1)$ transformation when $\hat w = -w$.

The modular covariant derivative,
\eqn\derdef{{\cal D}_w =i \left( {\partial \over \partial \tau} - i {w \over
2\tau_2}\right),}
maps  $F^{(w,\hat w)}$ into $F^{(w+2,\hat w)}$  while the
anti-holomorphic covariant derivative,
$\bar {\cal D}_{\hat w} =
{\cal D}^*_{\hat w}$,  maps  $F^{(w,\hat w)}$ into $F^{(w,\hat w+2)}$.
It is more convenient
for our purposes to define the covariant derivatives,
\eqn\dertdef{D_w = \tau_2 {\cal D} = i \left(\tau_2 {\partial \over
 \partial \tau}  - i {w \over
2} \right), \qquad \bar D_{\hat w} =
\tau_2 \bar {\cal D} = -i \left(\tau_2 {\partial \over
 \partial\bar  \tau}  + i {\hat w \over
2} \right) }
which change the $U(1)$ charge of $F$ by two units,
\eqn\derivprop{D_w F^{(w,\hat w)} = F^{(w+1,\hat w -1)}, \qquad
\bar D_{\hat w} F^{(w,\hat w)} = F^{(w-1,\hat w+1)}.}

The Laplace operator on the fundamental domain of $SL(2,\IZ)$ is defined to be,
\eqn\lapon{\nabla^2_0 \equiv
\nabla^2 = 4 \tau_2^2 {\partial \over \partial \tau } {\partial \over
 \partial \bar \tau} ,}
when acting on $(0,0)$ forms.  More generally, we shall be interested
in the Laplacian acting on $(w, - w)$ forms.   There are two such
Laplacians which are  defined by,
\eqn\lapthree{\nabla^2_{(-)w} = 4 D_{w-1} \bar
D_{-w} = 4 \tau_2^2
  {\partial \over \partial \tau } {\partial \over
 \partial \bar \tau} - 2 iw \left( {\partial \over  \partial \tau } + {\partial
\over  \partial \bar \tau}\right) -  w(w-1),}
and
\eqn\lapagain{\eqalign{\nabla^2_{(+) \, w} = &
 4 \bar D_{-w-1} D_w =  4 \tau_2^2
  {\partial \over \partial \tau } {\partial \over
 \partial \bar \tau} - 2 iw \left( {\partial \over  \partial \tau } +
 {\partial \over  \partial \bar \tau}\right) -  w(w + 1), \cr
 = & \nabla^2_{(-)w} - 2w .\cr}}

Now consider a $(w,-w)$ form that is an eigenfunction of the Laplace operator
$\nabla^2_{(-)\, w}$ with   eigenvalue $\sigma_w$,
\eqn\eigenf{\nabla^2_{(-)\, w}\, F^{(w,-w)}
= 4D_{w-1} \, \bar D_{-w} \, F^{(w,-w)} = \sigma_w \, F^{(w,-w)}.}
Applying $\bar D_{-w}$ to this equation gives,
\eqn\neweq{\nabla^2_{(+)w-1}\, F^{(w-1,-w+1)} = \sigma_w \,
F^{(w-1,-w+1)}.}
It is also easy to see that,
\eqn\eagers{\eqalign{\nabla^2_{(-)\, w-1}\, F^{(w-1,-w+1)} & =
  4D_{w-2} \, \bar D_{-w+1} \, F^{(w-1,-w+1)}, \cr &= (\sigma_w + 2w-2)
\, F^{(w-1,-w+1)},}}
where $F^{(w-1,-w+1)} = \bar D_{-w}\, F^{(w,-w)}$.  Repeating this for
$m$ steps gives
\eqn\eigenm{\eqalign{\nabla^2_{(-)\, w-m}\, F^{(w-m,-w+m)}
=& 4D_{w-m-1} \, \bar D_{-w+m+1} \, F^{(w-m,-w+m)}, \cr
=& (\sigma_w + 2mw-m^2-m) \, F^{(w-m,-w+m)}.\cr}}
Similarly,
\eqn\eigendu{\nabla^2_{(+)\, w-m}\, F^{(w-m,-w+m)}
= (\sigma_w + 2mw -2w - m^2 +m) \, F^{(w-m,-w+m)}.}
This relation between eigenvalue equations will be useful in analyzing
the  equations that are
satisfied by the modular forms that enter in $S^{(3)}$.

An indication of why this is so comes  from various
duality arguments that relate type
 II string theories and M theory.   Firstly,
it was argued in \refs{\greengut,\greengutvan} that
the function $f^{(0,0)}$ should satisfy \lapone, in which case it should be
 an eigenfunction of the $\nabla^2_0$  on the fundamental domain of
$SL(2,\IZ)$ with eigenvalue $3/4$.  Furthermore,  in \greengutkwon\ it was
argued  that the nonholomorphic
modular forms that arise as coefficients in $S^{(3)}$   are
 related to each other by applying covariant derivatives.  For
example, it was suggested that\foot{We are
using  a more uniform notation
 for the modular forms here than in \greengut.}
\eqn\twelvedef{f^{(12,-12)} = (\tau_2 {\cal D})^{12} f^{(0,0)}=
 D^{12} f^{(0,0)}\equiv D_{11} \cdots D_1 D_0 f^{(0,0)}   .}
Using this relation and \eigenm\ for the case $m=w$, and assuming that
 $f^{(0,0)}$ indeed satisfies \lapone, leads to the eigenvalue equation
that $f^{(12,-12)}$ is expected to satisfy,
\eqn\laptwelve{\nabla^2_{(-)\, 12} f^{(12,-12)} = \left( -132 +
{3\over 4}\right)  f^{(12,-12)}.}
In the next section we will prove, using supersymmetry alone, that
$f^{(12,-12)}$ does satisfy this equation.

The solution to the Laplace equation \lapone\ with eigenvalue $\sigma =3/4$
is unique if we assume that
 $f^{(0,0)}$ has a power law behavior near the boundary of the
fundamental domain of $SL(2,\IZ)$ which agrees with the known tree-level and 
one-loop contributions.  More generally, let us denote a  
solution of the scalar Laplace
equation  with eigenvalue $\sigma = s(s-1) >1/4$ by
 $E_s(\tau)$ \terras,
\eqn\eslap{\nabla^2 E_s = s(s-1) \, E_s.}
We can express $E_s (\tau)$ in terms of the nonholomorphic Eisenstein series,
\eqn\esdef{E_s (\tau) = {1\over 2} \tau_2^s \, \sum_{(m,n)=1}|m\tau
+n|^{-2s}, }
where $(m,n)$ denotes the greatest common divisor of $m$ and $n$. 
The eigenfunctions $E_s (\tau)$ are singled out by their power law behavior 
near the boundary of the moduli space, which agrees with the known tree-level 
and
perturbative contributions to the interactions that we are  considering.
 It follows from \twelvedef\ that 
$f^{(12,-12)}$ is also determined uniquely by its Laplace equation \laptwelve\
if the presence of a tree-level term is assumed.

While the arguments leading to \laptwelve\  were
motivated in prior work rather
indirectly by various dualities,
our purpose is to prove that \laptwelve\ and related conditions follow
directly from a rather simple application of supersymmetry.

\newsec{Determining Terms in $S^{(3)}$ Using Supersymmetry}

We now proceed to a precise determination of the modular forms that
enter into \nonpert.
This starts by selecting two specific terms in the effective
Lagrangian at order $(\alpha')^3$,
\eqn\starti{L^{(3)}_1 =  {\det\, e} \,  \left(
f^{(12,-12)}(\tau,\bar\tau) \lambda^{16} + f^{(11,-11)}(\tau,\bar\tau)
 \hat G \lambda^{14}\right). }
Our notation is chosen so that $(\alpha')^{n-4}\int d^{10} x L^{(n)} = 
S^{(n)}$.
We will show that these are related by a subset of the  supersymmetry
transformations that do not mix with any of the other terms at
this order. We will also take into account terms from the variation
of the lowest order action, $S^{(0)}$, that can mix with the variations
of \starti.

We will only need to consider those
terms in $\hat G$ that are bilinear in the fermions. After using the identity,
\eqn\identh{(\gamma^{\mu\nu\rho}\gamma^0)_{a
b}\, (\lambda)^{14}_{a b} \,  (\bar \psi_\mu
\gamma_{\nu\rho}\lambda) =  - 144 \bar \psi_{\mu}\, \gamma^{\mu}
\gamma^0 \,\lambda^{15} = 144 \lambda^{15} \gamma^\mu\psi_\mu^*,}
where we have used the fact that $\gamma_{\nu\rho}
\gamma^{\mu\nu\rho} = - 72 \gamma^\mu$,  the relevant terms
in $L^{(3)}_1$ can be expressed as
\eqn\relterm{L^{(3)}_1 =  {\det\,e}
\, \left( f^{(12,-12)} \lambda^{16} - 3
\cdot 144
f^{(11,-11)} \, \left(\lambda^{15} \gamma^\mu \psi_\mu^*\right) + \ldots
\right).}
The ellipsis represents other terms in $\hat G$ which do not
affect the subsequent argument.

First consider the lowest order supersymmetry transformation of
\relterm\  into  $\det \, e\, \lambda^{16}\,
\psi_\mu^*\,\epsilon$. From \relterm\ we have,
\eqn\varone{\eqalign{\delta_1^{(0)} L^{(3)}_1 =&  
\delta^{(0)}( \det\,e)  \,
f^{(12,-12)} \lambda^{16} +   {\det\, e}
\, f^{(12,-12)} \delta^{(0)}(\lambda^{16}) \cr
&  - 3\cdot  144\, {\det\,e}
\, \left( {\partial \over  \partial \tau}
f^{(11,-11)} \delta^{(0)} \tau\, \left(\lambda^{15}\gamma^\mu
\psi_\mu^*\right)   +
f^{(11,-11)} \delta^{(0)}\left(\lambda^{15}\,
\gamma^\mu \psi_\mu^*\right)\right)\cr
=& i\,  {\det \, e}
\, \left(\bar \epsilon^* \gamma^\mu \psi_\mu^*\,  f^{(12,-12)}
\lambda^{16} + {1\over 8} (\lambda^{15})_a
\left(\gamma^{\mu\nu\rho}\epsilon \right)_a \, \bar \psi_{[\mu}
\gamma_{\nu\rho]} \lambda f^{(12,-12)}\right. \cr
&\left. + 6 \cdot  144i\, D_{11} f^{(11,-11)} (\lambda^{15})_a
\left(\gamma^\mu\psi_\mu^*\right)_a \bar \epsilon^*\lambda\right) \cr
=&-i \,  {\det\,e}
\,  \left(\bar \epsilon^* \gamma^\mu \psi_\mu^*\right)\,  \lambda^{16}\,
\left( 8 f^{(12,-12)} + 6 \cdot  144\, D_{11} f^{(11,
-11)}\right),\cr}}
where we have only kept terms proportional to  $\lambda^{16}\,
\psi_\mu^*\,\epsilon$.  In passing from the first to the second line
in this equation, we have made use of the standard
$\delta^{(0)}$
supersymmetry transformations
 summarized in  Appendix A.  It is important to check whether there could
also be a contribution of the same form as \varone\ arising from a
$(\alpha')^3\delta^{(3)}$
variation of the fields  in the lowest order action $S^{(0)}$.
However, it is easy to see by inspection
that no term with $\lambda^{16} \psi^*_\mu$  can arise from the
variation of any term in $S^{(0)}$.\foot{The only relevant terms are
those involving only fermionic fields  since bosonic fields vary into
derivatives.  The only fermion interactions that could vary into the required
form would be terms such as $\lambda^2 \lambda^* \psi^*_\mu$, $\lambda^3
\psi^*_\mu$, $\ldots$, which are excluded from the classical theory
since they violate $U(1)$ charge conservation.}
  This means that we must require $\delta_1^{(0)} L^{(3)}_1 =0$,
which implies that
\eqn\constwo{D_{11} f^{(11,-11)} =  -{4\over 3 \cdot  144}
f^{(12,-12)}.}
This condition is consistent with the modular weights assigned to the
functions  $f^{(w,-w)}$.

In order to find another condition relating $f^{(12,-12)}$ and
$f^{(11,-11)}$, we  now consider the term in the variation of \relterm\
 that is proportional to  $ \det\, e\, \lambda^{16} \,
\lambda^*\,\epsilon^*$.  This term is,
 \eqn\vartwo{\eqalign{\delta_2^{(0)} L^{(3)}_1 =& {\det\, e }\,
\left(
{\partial f \over \partial \bar \tau}^{(12,-12)}\delta^{(0)} \bar \tau \,
  \lambda^{16} + f^{(12,-12)} \lambda^{15} \delta^{(0)} \lambda  -3
\cdot  144 f^{(11,-11)}
\lambda^{15}\delta^{(0)}(\gamma^\mu  \psi_\mu^*)\right)\cr
 = &- 2i \, {\det\, e}
\, \lambda^{16}  (\bar\epsilon \lambda^*)
\left(- i\left( \tau_2 {\partial \over \partial \bar \tau} -  6i \right)
 f^{(12,-12)} + 3\cdot 144\cdot {15\over 2} f^{(11,-11)}   \right) +
\ldots
\cr  =&  - 2i\, {\det\, e}
\, \lambda^{16}  (\bar\epsilon \lambda^*)
\left(\bar D_{-12}   f^{(12,-12)} +
3\cdot 144\cdot {15\over 2} f^{(11,-11)}   \right) + \ldots,
\cr}}
where we have made explicit
only the terms containing $\lambda^{16} \lambda^*\epsilon^*$
in the second line.

In this case, there is another contribution of the same form as 
 $\delta_2^{(0)} L^{(3)}_1$ that arises from the $(\alpha')^3 \delta^{(3)}$
variation of  terms
in the lowest order IIB  Lagrangian $L^{(0)}$ (recall that we are really
using the action as a shorthand for the IIB equations of motion).   Even
though the complete set of interactions in the classical theory is not
tabulated explicitly in the
literature (it is implicit in the superspace
 formulation \howest),
it is easy to convince oneself that
the only possible term that can vary into $\delta_2^{(0)} L^{(3)}_1$
is a term of the form,
\eqn\nonterm{L^{(0)}_1 = -{c\over 6} \, \det\, e\,
\bar \lambda \gamma^{\mu\nu\rho} \lambda^*\, \bar
 \lambda^*\gamma_{\mu\nu\rho}
 \lambda ,}
which is the unique tensor structure
containing $\lambda^2 \lambda^{*2}$.  The coefficient $c$ has been
left free in this formula, but it is determined by the lowest order
 supersymmetry transformations.  It is  determined
 in Appendix B  by considering the mixing of $L_1^{(0)}$ with
\eqn\termtwo{L_2^{(0)} = {3\over 2}i {\det\, e} \bar
\lambda\gamma^\mu \lambda\, Q_\mu,}
and
\eqn\termthree{L_3^{(0)} = i\, {\det\, e}\, \bar \lambda
\gamma^\mu\gamma^\omega \psi_\mu^*
 P_\omega.}
The term $L_2^{(0)}$ is the connection part of the Dirac action for
the dilatino,
\eqn\connterm{i \int d^{10} x\, \det\, e\, \bar\lambda \gamma\cdot D \lambda, }
and its coeffcient is determined by the $U(1)$ charge for $\lambda$.  The
normalization of the second term can be extracted from the gravitino
field equation  (eq. (4.12) of \schwarza), but is also determined  by the
supersymmetry considerations in Appendix B.
The value of $c$ deduced in Appendix B is
\eqn\cdet{c =- {3\over 128}.}
Of course, the arbitrary Newton coupling has been set equal to a particular
value in defining the absolute normalization of the action, but this value
cancels out of all that follows.

We can now see that $L_1^{(0)}$ can vary into the same form as
$\delta_2^{(0)} L_1^{(3)}$ if we assume a variation of $\lambda^*$ of the form,
\eqn\newvar{\delta^{(3)} \lambda^*_a =- {1\over 6}\,i
  g(\tau,\bar\tau)\,  (\lambda^{14})_{cd}
(\gamma^{\mu\nu\rho} \gamma^0)_{dc} \, (\gamma_{\mu\nu\rho} \epsilon^*)_a,}
where $g(\tau,\bar\tau)$ is a function to be determined.  We will show
momentarily
that there must be such a term in the variation of $\lambda^*$ if the
supersymmetry
algebra is to close. Substituting in \nonterm\
gives a contribution,
\eqn\threelone{\eqalign{
\delta^{(3)} L^{(0)}_1 & = {2 c\over 36}\, i\, 
 {\det\, e}\,
g(\tau,\bar\tau)\, \bar\lambda
\gamma^{\mu\nu\rho}\gamma_{\rho_1\rho_2\rho_3} \epsilon^*\,
(\lambda^{14})_{cd} (\gamma^{\rho_1\rho_2\rho_3} \gamma^0)_{dc}\, \bar
\lambda^* \gamma_{\mu\nu\rho} \lambda
 \cr
& =- 480 \cdot 16\, i\,   c \, {\det\, e} \, g(\tau,\bar\tau) \lambda^{16}
(\bar\epsilon \lambda^*).\cr}}
Some of these manipulations make use of the gamma matrix identities listed
in  Appendix (A.1).
Comparing with \vartwo\ we see that in
order for the total contribution to $\delta L_1$ to vanish at order
$(\alpha')^3$, there must be a linear relation between the function
$g$ and the functions $f^{(11,-11)}$ and $\bar D_{-12} f^{(12,-12)}$,
\eqn\linrel{\bar D_{-12} f^{(12,-12)} \, + 3\cdot 144 \cdot {15\over 2} \,
f^{(11,-11)} + 240\cdot 16 \, cg =0.}

A further constraint on these functions is obtained by requiring the
supersymmetry algebra to close which requires the use of the
fermionic equations of motion as well as the equation for $F_5$. In fact,
these equations of motion
were determined in the classical theory by requiring closure of the 
superalgebra  for the low-energy type IIB 
theory in \schwarza.   Another important feature of supergravity theories
such as this is that the algebra need only close up
to a field-dependent 
local symmetry transformation.  In the case at hand, the important
fact is that the commutator of two supersymmetry transformations,
$[\delta_1,\delta_2]$,  gives
the usual transport term, $\bar\epsilon_2\gamma^\mu \epsilon_2 D_\mu$,
together with a supersymmetry transformation, $\delta_{\hat \epsilon}$,
and terms that vanish by the
equations of motion. The supersymmetry parameter $\hat \epsilon$ is field
dependent.

We will consider closure of the supersymmetry transformations
on the field $\lambda^*$.  First,
keeping only the terms linear in $\lambda$ derivatives,  we find (as in
eq. (4.5) of \schwarza),
\eqn\susyclos{\eqalign{(\delta^{(0)}_1 \delta^{(0)}_2 -
\delta^{(0)}_2\delta^{(0)}_1) \lambda^* &=  \xi^\mu  D_\mu
 \lambda^* -{3\over 8} i [\bar\epsilon_2 \gamma^\rho \epsilon_1 -
 (1\leftrightarrow 2)] \gamma_\rho \gamma^\mu D_\mu \lambda^*\cr
&  -{1\over 96}i [\bar \epsilon_2 \gamma^{\rho_1\rho_2 \rho_3}
 \epsilon_1 - (1\leftrightarrow 2)] \gamma_{\rho_1\rho_2\rho_3}
 \gamma^\mu D_\mu \lambda^*,\cr}}
where,
\eqn\param{\xi^\mu = - 2\Im \bar\epsilon_2 \gamma^\mu \epsilon_1.}
The first term on the right-hand-side is of the form expected for the
commutator of two supersymmetry transformations.  The remaining
terms are proportional to
the lowest-order term in the $\lambda^*$ equation of motion.
Many other terms that we will not need also
contribute to the full commutator to complete the low-energy
$\lambda^*$  
field equation on the right-hand-side, as well as generating local
transformations of $\lambda^*$.

  The higher order terms in  $L^{(3)}$ modify the equations
of motion and this should also be apparent by considering the closure of
the algebra.  Therefore, we now consider
terms that enter at order $(\alpha')^3$ from the commutator of a $\delta^{(0)}$
with a $\delta^{(3)}$.  More precisely, we shall consider terms in the
commutator involving only $\epsilon_2^*$ and $\epsilon_1$,
\eqn\termone{\eqalign{\left( \delta^{(0)}_{\epsilon_1}
\delta^{(3)}_{\epsilon^*_2}  \right.  & \left. - \delta^{(3)}_{\epsilon_2^*}
\delta^{(0)}_{\epsilon_1}\right) \lambda^*_a
=  - {1\over 3}\left( \tau_2 {\partial  \over \partial \tau}
 - i {45 \over 8}  \right)i\, g \, (\bar \epsilon_1^*
\lambda)( \lambda^{14})_{cd}(\gamma^{\mu\nu\rho}\gamma^0)_{dc}
 (\gamma_{\mu\nu\rho}  \epsilon_2^*)_a \cr
=& {2\over 48}\cdot {8 \over 3}\cdot 288\lambda^{15}_b\,i\,  \left[{3\over 8}
 \bar\epsilon_2 \gamma^\mu \epsilon_1
  (\gamma_\mu)_{ba} + {1\over 96} \bar \epsilon_2
\gamma^{\mu\nu\rho} \epsilon_1 (\gamma_{\mu\nu\rho})_{ba}\right]
 \left( \tau_2{\partial  \over
\partial \tau} - i{45\over 8}\right)g\cr
=& 32\,  D_{11}g \, \lambda^{15}_b\,  \left[{3\over 8}
 \bar\epsilon_2 \gamma^\mu \epsilon_1
  (\gamma_\mu)_{ba} + {1\over 96} \bar \epsilon_2
\gamma^{\mu\nu\rho} \epsilon_1 (\gamma_{\mu\nu\rho})_{ba}\right] \, +
 \,  \delta_{\hat\epsilon} \lambda^* .\cr}}
In passing from the first to the second equation, we have used once more
the Fierz identity and various gamma matrix
identities given in Appendix (A.1).  In the last line, we have
separated a term,
\eqn\newgauge{ \delta_{\hat\epsilon} \lambda^* = -i{1\over 24}g \,
 (\bar \epsilon_1^*
\lambda)( \lambda^{14})_{cd}(\gamma^{\mu\nu\rho}\gamma^0)_{dc}
 (\gamma_{\mu\nu\rho}  \epsilon_2^*)_a,}
 which is to be identified with a
supersymmetry transformation of the form \newvar\ with a particular
field dependent coefficient,
\eqn\partcoeff{\hat \epsilon = {i\over 4} \epsilon_2^* \,
 (\bar\epsilon_1^* \lambda).}
   This is unambiguously identified by the
 fact that it is needed in order to change the  $45/8$ in the
 previous lines to the $44/8$ which is contained in $D_{11}$.  This  is
correlated with the
fact that the function $g$  transforms with weight $(11,-11)$.

In writing \termone, we have taken pains to express the 
right-hand-side as a sum of precisely the same tensor structures that appear
   on the right-hand-side of \susyclos. Combining \susyclos\ and
\termone\ 
(including the powers of $\alpha'$) we
see that in order for the right-hand side of the commutator to vanish
the $\lambda^*$ field equation must be of the form,
\eqn\propr{ i \gamma^\mu D_\mu \lambda^* - (\alpha')^3\,  32\, D_{11}g\,
 \lambda^{15} + \dots =0,}
where the ellipsis indicates terms with different structure that we
have not considered.
 This equation has to be identified with the appropriate sum of 
terms in the $\lambda^*$
 equation of motion that  is obtained by varying the action 
   with respect to $\lambda$.  At the same order in $\alpha'$ this is
 given by,    
\eqn\lammot{ i\gamma^\mu D_\mu \lambda^*  -  (\alpha')^3  
f^{(12,-12)}\, \lambda^{15} + \dots
=0,}
where we have  only made explicit the term that is proportional to
$\lambda^{15}$.
Comparing \propr\ and \lammot\ gives the relation,
\eqn\neweqm{32 D_{11} g =  f^{(12,-12)}.}

Substituting \neweqm\ into \constwo\ gives,
\eqn\grelf{ g=  - {3\cdot 144 \over 128}   f^{(11,-11)}.}
There is no ambiguity in this relation between $g$ and $f^{(11,-11)}$ because 
there is no solution to  $D_{11} g =0$.
Substituting \grelf\  into \linrel\ using the value $c= - 3/128$ gives,
\eqn\twelveg{\bar D_{-12} f^{(12,-12)} = 3\cdot 144 \left(-{15\over 2} +
{45\over 64}  \right) f^{(11,-11)}.}
The two simultaneous  first-order differential equations,
\twelveg\ and \constwo\
are simply reduced to the independent second-order equations,
\eqn\laplacee{\eqalign{ \nabla^2_{(-)\, 12}f^{(12,-12)} =
 4 D_{11} \bar D_{-12}  f^{(12,-12)}
=& \left(-  132 + {3\over 4}\right) f^{(12,-12)}\cr
\nabla_{(+)\, 11}^2 f^{(11,-11)} \equiv  4 \bar
 D_{-12} D_{11} f^{(11,-11)} =& \left(-  132 + {3\over 4}\right)
 f^{(11,-11)}
 .\cr}}
The first of these equations is the same as \laptwelve.  
Therefore, the  modular form $f^{(12,-12)}$ is uniquely determined to be
the function suggested in \greengutkwon\ if we assume that
$f^{(12,-12)}$ has a tree-level and one-loop contribution at 
weak coupling.  This function can be expressed as
$D^{12}  f^{(0,0)}$  where $f^{(0,0)}$ satisfies the Laplace equation
\laptwelve\ with eigenvalue $3/4$ (the proof 
 that this function is actually 
the coefficient of the $\calR^4$ term will follow from the argument 
in the next paragraph).  Similarly, the second equation in
 \laplacee\ gives a unique expression for the modular form $f^{(11,-11)}$.

Having  determined $f^{(12,-12)}$ and 
$f^{(11,-11)}$  we would now like  to determine the 
 remaining terms in \nonpert\ of the same order but lower $U(1)$ charge,
such as $\calR^4$.  A simple way to determine these terms is
to consider the constraints on the coefficient functions that follow
from linearized supersymmetry and then to impose the requirement that the 
effective action be $SL(2,\IZ)$ invariant.
Linearized supersymmetry,  described in section 2, is 
valid to leading order in $(\tau_2)^{-1}$.   We saw that 
in that approximation the 
terms in \nonpert\ are expressed as  
an integral of a function of the  superfield $F[ \Phi] $ over one-half of 
superspace. Furthermore, it was argued in \greengutkwon\ that 
the linearized approximation is exact for the leading charge $K$ 
D-instanton contributions to the coefficient 
functions, $f^{(p,-p)}$.  These  
can be extracted by choosing $F[\Phi] = e^{2\pi i K\Phi}$ and 
agree with the expectation that the coefficients are related by
\eqn\expect{ f^{(p,-p)} = D_{p-1} \cdots D_0 f^{(0,0)}. }
Only the abelian pieces of the covariant derivatives
affect the argument
 to leading order in $(\tau_2)^{-1}$ which does not build in the required 
modular invariance.   The  modular covariant expressions are reproduced 
by using the fully modular covariant derivatives in \expect.

It should, of course, also  be true that the expressions for 
all the coefficients, $f^{(p,-p)}$,  in \nonpert\ also emerge from a 
more detailed application of the Noether procedure that  
considers all the possible mixing of terms in $S^{(3)}$
with arbitrary $U(1)$ charges.

\newsec{Comments on Higher Derivative Interactions}

\subsec{Some general comments}

More speculative extensions of the $\calR^4$ conjecture have been suggested
in \refs{\russo, \berkovafa,\rstrings}.  
 For example, 
interactions of the form, 
 \eqn\conj{\eqalign{(\alpha')^{-4}
 \sum_{g, \hat g =1}^\infty \sum_{p= 2-2g}^{2g-2}& 
(\alpha')^{2g+2\hat g -1} \int d^{10}x\, \det\, e\, F_5^{4 \hat g-4} 
G^{2g-2+p} G^{*\, 2g -2 -p} \, \cr
& (f_{g+ \hat g-1}^{(p,-p)}(\tau,\bar \tau) \calR^4 + \dots + 
f_{g+ \hat g-1}^{(12+p, -12-p)}(\tau,\bar \tau) \lambda^{16}),\cr}}
arose in \refs{\berkovafa,\rstrings}.\foot{More precisely, 
the interactions suggested in \refs{\berkovafa,\rstrings} only included the 
$\calR^4$ terms in this expression.}  The case $g=\hat g = 1$ 
corresponds to the terms that we considered in the earlier sections.
The modular functions $f_g^{(q,-q)}$ 
are expected to be given by the  generalized Eisenstein series,
\eqn\geneisen{f^{(q,-q)}_g = \sum_{(m,n) \ne (0,0)}{\tau_2^{g+\half}\over 
(m+n\tau)^{g+\half +q}(m+n\bar\tau)^{g+\half-q}}.}
Note that for $q=0$, these coefficient functions are proportional to $\ E_{g+ 
\half} 
(\tau)$, where $E_s$ was defined in \esdef.  Expanding  \geneisen\ for small 
coupling ($\tau_2 \to \infty$) leads, 
as in the case $g=1$, to two power-behaved terms that are to be identified with
perturbative terms in string theory.   These correspond to a tree-level term
and a  $g$-loop term.  In fact, the case  $s= {3\over 2}$ is the 
physical lower bound on $s$
since in that case  the loop term is of the lowest possible genus,
$g=1$.    The agreement of the perturbative behavior of \geneisen\
with the known perturbative contributions to \conj\ computed in \rvb\ 
is a primary motivation for the form of these coefficient functions. 
The perturbative
contributions were computed in a topological formalism further studied 
in \rvo. As in 
the case $g=1$, there are no higher order 
perturbative corrections but there is an infinite series of D-instanton 
corrections.  
 The conjectured functions  $f_g^{(q,-q)}$ in  \geneisen\ are 
again  eigenfunctions of the Laplace operator acting on $(q,-q)$ forms, as in 
the $g=1$ case. 
Now, however, the eigenvalue depends on $g$.  For example,
\eqn\fourr{4 \tau_2^2 \, \partial_\tau \partial_{\bar \tau} f^{(0,0)}_g
 = \left({1\over 4} + {g\over 2} \right) f^{(0,0)}_g.} 

{}From the perspective of superspace, the status of terms with $g+\hat g >3$  
is quite different from the terms we considered in 
section 3 for which $g+\hat g =2$.  Those terms could be written as integrals 
over
$1/2$ the on-shell superspace, which is described in terms of a superspace with 
a 
single Weyl $SO(9,1)$ spinor.  For this reason, we could have anticipated
the fact that they satisfied very constraining nonrenormalization conditions.  
Cases in which  $g+\hat g = 3$ (terms of order $(\alpha')^5$ relative to 
the Einstein--Hilbert term) appear to be  
similarly special since, by
dimensional analysis, they correspond to integrals over $3/4$ of the on-shell
superspace, i.e., over 24 Grassmann spinor components.  Since there is no
covariant description of $SO(9,1)$ spinors with 24 components, there is no
obviously simple superspace description of such terms.  However, as we will see 
in the next subsection an analysis of the supersymmetry transformations 
similar to the preceding one is likely to 
determine the form of these  $O((\alpha')^5)$ terms and provide further
motivation for the conjectured terms in \conj\ at this order.

\subsec{An outline of how  terms in $S^{(5)}$ are constrained}

We will  not  present a detailed analysis of
terms in  $S^{(5)}$ but rather, we will give a schematic outline of
how supersymmetry constrains at least some of 
these terms.  Consequently, we will not be concerned  about the 
exact normalizations or tensor structures that arise in  the
various terms.

We will consider interactions  in  $S^{(5)}$
with ${\hat g}=1$ and $g=2$, which  are terms of order  
$(\alpha')^5$ relative to the 
Einstein--Hilbert term.  An important consideration is that the absence of 
$(\alpha')$ and $(\alpha')^{2}$ 
corrections to the effective action (the absence of $S^{(1)}$  and $S^{(2)}$ 
terms) means that the
supersymmetry transformations have modifications that begin with $(\alpha')^3
\delta^{(3)}$.  These transformations do 
not mix any of the lower order terms 
in  $S^{(0)} + S^{(3)}$ with the terms in $S^{(5)}$. We therefore only 
need to consider $S^{(0)} + (\alpha')^5 S^{(5)}$ and $\delta^{(0)} +  
(\alpha')^5\delta^{(5)}$. 

In complete analogy to our earlier analysis, we will begin by considering
the term in $L^{(5)}$ of modular weight $(14,-14)$, 
\eqn\mone{L^{(5)}_1 = \det \, e\, \lambda^{16} \, \hat G^4\, 
f_2^{(14,-14)}(\tau,\bar\tau),}
recalling that $\hat G$ is the supercovariant extension of $G$ containing 
fermion bilinears.  The tensor structure is hidden in the 
abbreviation $ \hat{G}^4$ which should read,
\eqn\tense{ t_{\mu_1\cdots \mu_{12}}{\hat G}^{\mu_1\mu_2 \mu_{3}} \cdots
{\hat G}^{ \mu_{10}\mu_{11} 
\mu_{12}}, }
for a tensor structure $t$ which we will not specify here but would be
determined in a more complete treatment. 

As before, the first supersymmetry variation of \mone\ to consider is  the one
acting on $\bar\tau$ given in (A.21),  
\eqn\varyo{\delta^{(0)}_1 L^{(5)}_1 = - 2 \det \, e \, 
\lambda^{16} \,(\bar\e \lambda^*)
\hat G^4 \left(\tau_2 {\partial \over \partial \bar \tau} - 7i\right) 
f_2^{(14,-14)} (\tau,\bar\tau).}
In this case, there are two other 
 terms in $S^{(5)}$ that can vary into \varyo. 
The first is similar in structure to the term that appeared in our 
earlier analysis,
\eqn\mtwo{L^{(5)}_2 = \det \, e\,
  \lambda^{15} \gamma^\mu \psi^*_\mu\, \hat G^4\, f_2^{(13,-13)}
(\tau,\bar\tau),}
which is a piece of the supercovariant combination $  \det \, e 
\lambda^{14} \hat G^5 $. The relevant supersymmetry variation gives, 
\eqn\varyotwo{\delta_1^{(0)}L^{(5)}_2 = \det\, e \,\lambda^{15} \gamma^\mu 
\delta^{(0)}(\psi_\mu^*)\, \hat G^4\, f^{(13,-13)}(\tau,\bar\tau),}
where  $\delta^{(0)}(\gamma^\mu \psi_\mu^*)$ is given in Appendix A.

The second term is a new possibility, 
\eqn\mthree{L^{(5)}_3 =  \det \, e\, \lambda^{16} \, \hat G^3 \hat G^*\, 
\tilde f_2^{(13,-13)}(\tau,\bar\tau).}
The relevant part of this expression is the fermion bilinear in $\hat G^*$
proportional to $\psi \lambda^*$. Since   $\delta_1^{(0)}\psi $ contains
a $ \hat G \e^* $ piece, the variation
\eqn\varythree{\delta_1^{(0)} L^{(5)}_3 =  \det \, e\, 
\lambda^{16} \, \hat G^3 (\delta_1^{(0)} \hat G^*)\, 
\tilde f_2^{(13,-13)}(\tau,\bar\tau),}
 mixes with \varyo.

In addition, it is necessary to consider the mixing of these terms with 
terms of the 
classical action.  The two terms that are relevant are
 $L_1^{(0)}$ given in \nonterm\ and $L_4^{(0)}$  given by, 
\eqn\lfour{L_4^{(0)} = \psi_\mu \gamma_{\nu\rho}\bar\lambda \, 
G^{\mu\nu\rho}.}
For these terms to mix with \varyo\ there need to be modifications to the 
supersymmetry transformations that take the schematic form,
\eqn\modific{ \eqalign{ \delta^{(5)} \lambda^* & \sim g_1(\tau, \bar\tau)
 \hat G^4  (\lambda^{14})_{cd}
(\gamma^{\mu\nu\rho} \gamma^0)_{dc} \, (\gamma_{\mu\nu\rho} \epsilon^*), \cr
    \delta^{(5)} \psi_\mu & \sim g_2(\tau, \bar\tau)
\lambda^{16} (\hat G^3 \e^*)_\mu. \cr    } }
Invariance under supersymmetry then gives a linear relation between the
functions, 
\eqn\neweqs{\bar D_{-14} f_2^{(14,-14)}, \quad f_2^{(13,-13)}, \quad 
\tilde f_2^{(13,-13)}, \quad g_1,\quad g_2.}

Additional constraints that relate $f_2^{(14,-14)}$ and 
$f_2^{(13,-13)}$ can be obtained  by considering a second 
supersymmetry variation that mixes $L^{(5)}_1$ and $L^{(5)}_2$ and with
no other 
terms at 
order $(\alpha')^5$. An appropriate transformation to consider is 
\eqn\varyoon{\eqalign{ \delta_2^{(0)} L^{(5)}_1 & = \delta_2^{(0)} 
(\det \, e\, \lambda^{16} \, 
 \hat G^4) f_2^{(14,-14)}, \cr
& \sim (\det \, e\, \lambda^{16} \, 
 \hat G^4) \bar \epsilon^*
\gamma^\mu \psi^*_\mu \, f_2^{(14,-14)} + \ldots, \cr} }
and
\eqn\varytwoo{\delta_2^{(0)} L^{(5)}_2 = 2 \det \, e\,\left(\tau_2 {\partial 
\over 
\partial
\tau} + { 13 \over 2}
i \right)f^{(13,-13)}_2 \, \lambda^{16} \, \bar \epsilon^*
\gamma^\mu \psi^*_\mu \, \hat G^4,}
where we are using parts of $\delta^{(0)}\lambda$ from (A.24), 
$\delta^{(0)} \, e^m_{\ \mu}$ from (A.23) and $\delta^{(0)}\, \tau$ 
from (A.21).
In addition we must consider the variation of a term $L_5^{(0)}$ in $S^{(0)}$
where $L_5^{(0)}$ takes the form,
\eqn\lfivedef{L_5^{(0)} = \bar \psi_\mu \gamma_\nu \psi^*_\rho\, 
G^{\mu\nu\rho}.}
A variation of this term which mixes 
with \varyoon\ and \varytwoo\  is induced by 
the new transformation,
\eqn\fivemope{\delta^{(5)} \psi_\mu^* = g_3(\tau,\bar\tau) \, \lambda^{16}
\, \hat G^3_{\mu\nu\rho}\,\gamma^{\nu\rho} \epsilon,}
where $g_3$ is another function that has to be determined. Invariance
under supersymmetry then relates $D_{13}f_2^{(13,-13)}, \, f_2^{(14,-14)}$
and $g_3$.   

The final set of constraints  follow from closure of the 
supersymmetry algebra on $ \lambda^*, \psi $ and $\psi^*$. The
part of the commutators,
\eqn\clost{[\delta_{\epsilon_1},\delta_{\epsilon_2^*}] \lambda^*, \qquad 
[\delta_{\epsilon_1},\delta_{\epsilon_2^*}] \psi_\mu, \qquad
 [\delta_{\epsilon_1^*},\delta_{\epsilon_2}] \psi_\mu^*,}
proportional to $(\alpha')^5$ gives  a sufficient number of relations
to determine $g_1, g_2$ and $g_3$ in terms of the coefficient functions 
in $S^{(5)}$.  For example, identifying the right-hand-side of 
 the commutator,
\eqn\comms{ \eqalign{ \com{\delta_{\e_1}}{ \delta_{\e_2^*}} \lambda^*  
& \sim \delta_{\epsilon_1}^{(0)}\left( g_1(\tau, \bar\tau)
\hat G^4 (\lambda^{14})_{cd}\right)
(\gamma^{\mu\nu\rho} \gamma^0)_{dc} \, (\gamma_{\mu\nu\rho} \e_2^*)
+\ldots, \cr 
& \sim D_{13}\, g_1\, \e_1\, \lambda^{15}\, \hat G^4\, \e^*_2 +
  g_1\, \e_1\, \hat G^*\, \lambda^{15}\, \hat G^3   \e^*_2
\ldots,} }
with the $\lambda^*$ equation of motion 
will allow us to relate $D_{13} \, g_1$ and $f_2^{(14,-14)}$ as well as
$g_1$ and $\tilde f_2^{(13,-13)}$, by analogy with  
 the case we studied earlier. As with the
earlier case, it is important to  also subtract  the variation in the
reverse order, $ \delta^{(5)}_{\e_2^*}\delta^{(0)}_{\e_1} \lambda^*$.
But we also need to add the  variations,
$(\delta^{(5)}_{\e_1} \delta^{(0)}_{\e_2^*} - \delta^{(0)}_{\e_2^*}
\delta^{(5)}_{\e_1})\lambda^*$,   which
give a  non-vanishing contribution to \comms\ although there was no
analogous contribution in the case considered in section 3. 
Such terms have been suppressed on the right-hand-side of 
 \comms\  but they will give 
additional contributions that must be taken  into account.
Likewise, 
the $(\alpha')^5$ part
of the commutator,
\eqn\eqsbv{\eqalign{ \com{\delta_{\e_1}}{ \delta_{\e_2^*}} \psi_\mu  & \sim
g_2(\tau, \bar\tau) \delta^{(0)}
\lambda^{16}  ( \hat G^3 \e^*_2 )_\mu + \ldots, \cr
& \sim  g_2  ( \e_1 \lambda^{15} \hat G^4   \e^*_2 )_\mu + \ldots,\cr}}
determines the $\psi_\mu$ equation of motion and
 relates $g_2$ to  $f_2^{(14,-14)}$.
Lastly,  $g_3$ is constrained by considering, 
\eqn\newsa{\eqalign{  \com{\delta_{\e_1^*}}{ \delta_{\e_2}} \psi^*_\mu  & \sim
\delta^{(0)} \left(
g_3(\tau, \bar\tau) \lambda^{16} (\hat G^3 \e_2)_\mu \right) + \ldots, \cr 
& \sim  \left(\bar D_{-13} g_3\right)\,   \e_1^*\, \lambda^*\, \lambda^{16}  
(\hat G^3 \e_2)_\mu + 
\ldots, \cr}}
which determines the $\psi^*_\mu$ equation of motion and 
relates $g_3$ and $\tilde  f_2^{(13,-13)}$.  In writing 
\eqsbv\ and \newsa\ we have again been symbolic and 
 suppressed the fact that it is
essential to include all the terms involving products of $\delta^{(0)}$ with
$\delta^{(5)}$ in the commutators, as
with \comms.

 The arguments of this subsection  demonstrate how
closure of the supersymmetry algebra
together with a judicious choice of supersymmetry variations of the Lagrangian
can completely determine the interactions in $S^{(5)}$.

\subsec{Future directions}

It is less clear how things might work for higher derivative terms in the
string effective action. 
The most significant new feature, which follows simply from
dimensional analysis, is that  
terms  in \conj\ that contribute to $S^{(7)}$ 
can arise from integration
over the whole of the superspace. We would not generally expect these terms to 
be protected. More pragmatically, at this order the Noether procedure
escalates in complexity.  This is largely because at order $p$, there are
many possible terms $ \delta^{(n)} S^{(m)}$ where $n+m=p$, that can mix
under supersymmetry.  

In the case of $p=7$, for example, $ \delta^{(4)} S^{(3)}$ can mix with 
$ \delta^{(7)} S^{(0)}$
and  $\delta^{(0)} S^{(7)} $. This kind of mixing certainly complicates 
the systematics at higher orders.
Nevertheless, it could still be the case that the conjectures in
\refs{\russo,\berkovafa,\rstrings} are correct. At least the terms
in \berkovafa\ were special in perturbative string theory because of their
relation to topological amplitudes, and this could be reflected in the 
systematics of the Noether construction. Should these conjectures prove true, 
they would point to some interesting and powerful 
implications of supersymmetry that would be satisfying to understand more
deeply.

Another avenue that would be very fruitful to explore is the 
generalization of this analysis to compactified supergravity.
  The simplest example
is the nine-dimensional theory with moduli space $SL(2,\IZ)\backslash 
SL(2,\IR)/O(2)
\times \IR$.  This 
can be viewed as M theory on a two-torus where the $SL(2,\IZ)$ acts on
the complex structure of the torus, $\Omega$, 
and $\IR$ is its volume, $V$.  The expected
$\calR^4$ term,  given in \greenvan, is of the form 
$(V^{-1/2}f^{(0,0)}(\Omega,\bar \Omega) + 2\pi^2/3 V)\calR^4$.  
 New features  enter the effective action in this case that are absent
at the boundary of moduli space corresponding to ten-dimensional type IIB 
theory.
Notably, the toroidal compactification of the eleven-form of \firsd\ enters
 the action.  An  indirect argument given in
\greenvan\ relates this by supersymmetry to the $\calR^4$ term but it should
now be possible to relate these terms directly.
It has been suggested
that in compactifications to lower dimensions,  the appropriate modular 
functions are those associated with eigenfunctions of the Laplace operator
on the U-duality moduli spaces \refs{\rkirit}.  These are cases that can 
certainly
be analyzed with the tools that we have developed here. It would be extremely
interesting to see what happens in low dimensions, where the U-duality
group becomes exceptional, and for sufficiently low dimensions, 
infinite-dimensional. These same techniques are also applicable to cases
with less supersymmetry. For example, 
compactifications of M theory on hyper\kh\
spaces, and toroidal compactifications of the heterotic or type I strings.  
 Undoubtedly, supersymmetry will continue
to yield new insights about the non-perturbative structure of string theory
and about M theory.

\vskip 0.5cm

\bigbreak\bigskip\bigskip\centerline{{\bf Acknowledgements}}\nobreak

This work was carried out during the  Amsterdam Workshop on
String Dualities and we are very grateful to the organizers for their
hospitality. It is also our pleasure to thank Michael Gutperle, 
Hirosi Ooguri and Mark
Stern for useful conversations, and the cafe de Jaren for its amenable
ambiance. The work of  S.S. is supported by
NSF grant DMS--9627351.

\vfill\eject

\appendix{A}{Type IIB Supergravity Revisited}

\subsec{ Some spinor and gamma matrix identities}

The spinors that enter into the IIB theory are complex Weyl spinors.
The gravitino and dilatino have opposite chiralities and the
supersymmetry parameter has the same chirality as the gravitino.  The
complex conjugate of the product of a pair of spinors is defined by
\eqn\compconj{ (\lambda_a\, \rho_b)^* = - \lambda_a^*\, \rho_b^*.}
The conjugate of any spinor is defined by,
$\bar \lambda = \lambda^* \gamma^0$.
We will choose our metric to be space-like and the $\gamma$ matrices to be real
and satisfy the Clifford algebra,
\eqn\cliffalg{ \left\{ \gamma^\mu, \gamma^\nu \right\} = 2 \eta^{\mu\nu}.}
Noting that, 
\eqn\gammzer{\gamma^0 \gamma^\mu = - (\gamma^\mu)^T \gamma^0,}
it follows that 
two complex  chiral spinors of the same chirality, $\lambda_1$ and
$\lambda_2$,  satisfy the relations,
\eqn\lamprops{\eqalign{\bar\lambda_1\gamma^\mu \lambda_2
= & - \bar\lambda_2^* \gamma^\mu
\lambda_1^*, \cr
\bar\lambda_1\gamma^{\mu\nu\rho} \lambda_2 =&
\, \bar\lambda_2^* \gamma^{\mu\nu\rho} \lambda_1^*, \cr
\bar\lambda_1\gamma^{\rho_1\dots \rho_5} \lambda_2 =&
 - \bar\lambda_2^* \gamma^{\rho_1 \dots \rho_5} \lambda_1^*,\cr}}
while two chiral spinors of opposite chiralities, $\lambda$ and $\epsilon$,
satisfy,
\eqn\newlams{\eqalign
{\bar\lambda \, \epsilon =&\, \bar\epsilon^*\, \lambda^*,\cr
\bar\lambda \gamma^{\rho_1\rho_2} \epsilon =&  -\bar\epsilon^*
\gamma^{\rho_1\rho_2} \lambda^*,\cr
\bar\lambda \gamma^{\rho_1\rho_2\rho_3\rho_4} \epsilon =&
\bar\epsilon^*\gamma^{\rho_1\rho_2\rho_3\rho_4} \lambda^*.
\cr}}
The Fierz identity for ten-dimensional complex Weyl spinors can be
expressed as,
\eqn\fierz{\lambda_1^{a} \bar \lambda_2^{ b} = -{1\over 16} \bar \lambda_2
\gamma_\mu \lambda_1\, \gamma^\mu_{ab} + {1\over 96} \bar \lambda_2
\gamma_{\mu\nu\rho} \lambda_1\,  \gamma^{\mu\nu\rho}_{ab} - {1\over 3840}
\bar\lambda_2 \gamma_{\rho_1 \cdots\rho_5}\lambda_1\,  \gamma^{\rho_1\cdots
\rho_5}_{ab},}
where $\lambda_1$ and $\lambda_2$ are two chiral spinors of the same chirality.

An additional useful identity is,
\eqn\lamidens{\gamma^{\rho_1\dots\rho_5}\lambda_1\,
\bar\lambda_2\gamma_{\rho_1\dots \rho_5} \lambda_3 =0,}
where $\lambda_1$, $\lambda_2$ and $\lambda_3$ are three chiral spinors of the
same chirality.

Some gamma matrix identities that are useful in proving the various
relationships in the text are,
\eqn\appone{\eqalign{\tr(\gamma_{\mu\nu\rho} \gamma^{\rho_1\rho_2\rho_3}) =&
-16
\left(\delta^{\rho_1}_\mu \delta^{\rho_2}_\nu\delta^{\rho_3}_\rho
- \delta^{\rho_2}_\mu \delta^{\rho_1}_\nu\delta^{\rho_3}_\rho
+ \delta^{\rho_2}_\mu \delta^{\rho_3}_\nu\delta^{\rho_1}_\rho \right. \cr
& \left. -\delta^{\rho_3}_\mu \delta^{\rho_2}_\nu\delta^{\rho_1}_\rho
+\delta^{\rho_3}_\mu \delta^{\rho_1}_\nu\delta^{\rho_2}_\rho
-\delta^{\rho_1}_\mu \delta^{\rho_3}_\nu\delta^{\rho_2}_\rho
\right).\cr}}
\eqn\appfive{\eqalign{\gamma^\mu\, \gamma_\sigma\, \gamma_\mu =& -8
\gamma_\sigma,\cr
\gamma^\mu \,\gamma_{\sigma_1\sigma_2\sigma_3}\,
\gamma_\mu =&  -4 \gamma_{\sigma_1\sigma_2\sigma_3},\cr
\gamma^\mu
\gamma_{\sigma_1\dots\sigma_5}
  \gamma_\mu = & \, 0,\cr
\gamma^{\mu\nu\rho} \, \gamma_\sigma\, \gamma_{\mu\nu\rho}
 =&   - 288 \gamma_{\sigma},\cr
\gamma^{\mu\nu\rho} \,
\gamma_{\sigma_1\sigma_2\sigma_3}\, \gamma_{\mu\nu\rho}
 =&  - 48 \gamma_{\sigma_1\sigma_2\sigma_3},\cr
\gamma^{\mu\nu\rho} \,
\gamma_{\sigma_1\dots\sigma_5}\, \gamma_{\mu\nu\rho}
 = & -14\gamma_{\sigma_1\dots\sigma_5} .\cr}}

\subsec{The fields and their supersymmetry transformations}

Here we will review various features of type IIB supergravity that
are useful in the body of the paper.  Most of this material can be found in
\schwarza\ in a form that is adapted to the field definitions
in which the global symmetry is $SU(1,1)$ and the scalar fields
parameterize the coset space $SU(1,1)/U(1)$, which is the Poincar\'e
disk.  It is simple to
transform this to our parameterization in which the global symmetry is
$SL(2,\IR)$
and the scalars parameterize the coset space $SL(2,\IR)/U(1)$,
or the upper half plane.

The theory is then defined in terms of the following  fields:
the scalar fields can be parameterized by the frame field,
\eqn\vdef{V \equiv \pmatrix{V^1_- & V^1_+ \cr
                            V^2_- & V^2_+ \cr} = {1\over \sqrt{-2i\tau_2}}
\pmatrix{\bar \tau e^{-i\phi} & \tau e^{i\phi} \cr
                e^{-i\phi} & e^{i\phi} \cr},}
where $V^\alpha_\pm$ ($\alpha =1,2$) is a $SL(2,\IZ)$ matrix that
 transforms from the left by the global
$SL(2,\IR)$ and from the right by the local $U(1)$. Note that we are using a
complex basis for convenience.
A general transformation is then written as,
\eqn\vtrans{(V_+^\alpha,V_-^\alpha)  \to U^\alpha_{\ \beta}
\left( V^\beta_+ e^{i\Sigma},  V^\beta_- e^{-i\Sigma}\right),}
where $U$ is a  $SL(2,\IR)$ matrix and $\Sigma$ is the $U(1)$ phase.
 An appropriate choice of $\Sigma$ fixes the gauge and
eliminates the scalar field $\phi$.  We will make the gauge  choice
 $\phi=0$.   Since this
 gauge is not maintained by generic symmetry  transformations,
it is necessary to compensate a symmetry transformation with an appropriate
local
 $U(1)$ tranformation to maintain the gauge.   In particular, the
 local supersymmetry transformations require compensating  local
 $U(1)$ transformations.  The supersymmetry and $U(1)$
transformations of $V_-^\alpha$ are given by,
\eqn\susyv{\delta^{(0)} V_-^\alpha = i V_+^\alpha \bar \epsilon \lambda^* - i
\Sigma V_-^\alpha.}
This choice ensures that
the gauge $\phi=0$ is maintained if a local supersymmetry transformation is
accompanied by a $U(1)$ transformation with parameter,
\eqn\sigtrans{\Sigma = {1\over 2} (\bar \epsilon \lambda^* - \bar\epsilon^*
\lambda).}

The $SL(2,\IR)$ singlet expression,
\eqn\qdef{Q_\mu =-i \epsilon_{\alpha\beta} \, V^\alpha_+
\partial_\mu V^\beta_- , }
is the composite $U(1)$ connection and transforms as $Q \to Q + \partial_\mu
\Sigma$ under infinitesimal local $U(1)$ transformations, while the
$SL(2,\IR)$
singlet expression
\eqn\pdef{P_\mu = - \epsilon_{\alpha\beta}V_+^\alpha \partial_\mu V^\beta_-,}
 transforms with $U(1)$ charge $q_{P}= 2$.  In the gauge $\phi=0$, the
expression
for $P_\mu$ takes the simple form,
\eqn\pfix{P_\mu = {i\over 2} {\partial_\mu \tau \over \tau_2}.}

The fermions comprise the complex chiral gravitino, $\psi^a_\mu$, which has
$U(1)$ charge $q_\psi=1/2$, and the
 dilatino, $\lambda^a$, with $U(1)$ charge $q_\lambda = 3/2$.  These
 two fields
 have opposite chiralities.
The graviton is a $U(1)$ and $SL(2,\IR)$ singlet as is the antisymmetric
fourth-rank potential, $C^{(4)}$,
 which has a field strength $F_5 = d C^{(4)}$.  As is well known, this field
strength has an equation of motion that is expressed by the self-duality
condition $F_5 = * F_5$, which cannot be obtained
from a globally well-defined Lagrangian.  For this reason, our
considerations are restricted to statements concerning the on-shell properties
of the theory where the
fields satisfy the equations of motion.

The two antisymmetric second-rank potentials,  $B_{\mu\nu}$ and
$C^{(2)}_{\mu\nu}$, have   field strengths $F^1$ ($NS\otimes NS$)  and $F^2$
($R\otimes R$)
that form an  $SL(2,\IR)$ doublet, $F^\alpha$.  It is very natural to package
them
into the $SL(2,\IR)$ singlet fields,
\eqn\gdef{G = - \epsilon_{\alpha\beta} V_+^\alpha F^\beta ,
\qquad G^* =-\epsilon_{\alpha\beta} V_-^\alpha F^\beta,}
which carry $U(1)$ charges $q_G =+1$ and $q_{G^*}= -1$, respectively.

In a fixed $U(1)$  gauge,  a global $SL(2,\IR)$ transformation which acts on
$\tau$ by
\eqn\slact{\tau \to {a\tau + b\over c\tau + d}, }
with $ad-bc=1$, induces
a $U(1)$ transformation on the fields that depends on their charge.
Thus, a field $\Phi$ with $U(1)$ charge $q_\Phi$ transforms as,
\eqn\uonep{\Phi \to \Phi \, \left({c\bar \tau + d \over c\tau +
d}\right)^{q_\Phi/2}.}
The higher derivative terms of interest to us only repect the
$SL(2,\IZ)$ subgroup of $SL(2,\IR)$ for which $a,b,c,d$ are integers and
the continuous $U(1)$ symmetry is broken.

The supersymmetry of the action is naturally described in terms of combinations
of bosonic fields and fermion bilinears which are \lq supercovariant', 
which means that they 
do not contain derivatives of the supersymmetry parameter $\epsilon$ in
their transformations.  These combinations are,
\eqn\supcov{\eqalign{ \hat G_{\mu\nu\rho}& = G_{\mu\nu\rho} - 3 \bar\psi
_{[\mu}\gamma_{\nu\rho]}\lambda - 6i \bar\psi^*_{[\mu} \gamma_\nu \psi_{\rho]},
\cr
\hat P_\mu &= P_\mu - \bar \psi^* \lambda, \cr
\hat F_{5\, \mu_1\dots,\mu_5} & = F_{5\, \mu_1\dots,\mu_5}
-5 \bar\psi_{[\mu_1} \gamma_{\mu_2\mu_3\mu_4} \psi_{\mu_5]}-
{1\over 16} \bar \lambda \gamma_{\mu_1 \dots \mu_5} \lambda   .\cr}}

We will now present the lowest-order supersymmetry transformations, suitably
adapted from those given in \schwarza\ to the $SL(2,\IR)$
parameterization. From \susyv\ and \sigtrans, it follows that
\eqn\tautrans{\delta^{(0)} \tau = 2\tau_2\bar \epsilon^* \lambda,
\qquad \delta^{(0)}\bar \tau = -2\tau_2\bar \epsilon \lambda^*.}
It follows from the definition of $Q_\mu$ and the transformations of
$\tau$ and $\bar \tau$ that
\eqn\qtrans{\delta^{(0)} Q_\mu = - \bar\epsilon \lambda^* \, P_\mu \,
+ \ c.c.}
Also, the supersymmetry transformation of the zehnbein is given by,
\eqn\zehntrans{\delta^{(0)} e^m_\mu = i(\bar \epsilon \gamma^m \psi_\mu +
\bar\epsilon^* \gamma^m \psi_\mu^*).}
The transformation of the dilatino is given, in the fixed $U(1)$
gauge, by
\eqn\lamtran{\eqalign{\delta^{(0)} \lambda =& i\gamma^\mu
\epsilon^* \,\hat P_\mu - {1\over 24} i \gamma^{\mu\nu\rho}\epsilon
\hat G_{\mu\nu\rho} + \delta^{(0)}_\Sigma \lambda\cr
= & i\gamma^\mu\epsilon^*\,\hat P_\mu +{i\over 8}  \gamma^{\mu\nu\tau}
\epsilon  \, \left(\bar\psi_{[\mu}\gamma_{\nu\tau]}
\lambda\right)  - i\gamma^\mu \epsilon^*\,
(\bar \psi^* \lambda) + \delta^{(0)}_\Sigma \lambda + \dots ,\cr}}
where we have only kept the terms that are needed in the body of this paper in
the second line. The $\delta_\Sigma$ arises from the compensating $U(1)$
gauge transformation,
\eqn\compenlam{\delta^{(0)}_\Sigma \lambda_a = {3\over 2}i \Sigma\, \lambda_a
=
{3\over 4} i\lambda_a
(\bar\epsilon \lambda^*)  - {3\over 4} i
\lambda_a (\bar\epsilon^* \lambda) .}
The gravitino transformation is given by,
\eqn\vartino{\eqalign{\delta^{(0)} \psi_\mu =& D_\mu \epsilon +
{1\over 480}i \gamma^{\rho_1\dots\rho_5}\gamma_\mu \epsilon \hat F_{\rho_1\dots
\rho_5}+ {1\over 96}\left(\gamma_\mu^{\ \nu\rho\lambda}\hat G_{\nu\rho\lambda}
-
9 \gamma^{\rho\lambda} \hat G_{\mu\rho\lambda}\right)\epsilon^*  \cr
& -{7\over 16} \left(\gamma_\rho \lambda\, \bar \psi_\mu \gamma^\rho \epsilon^*
- {1\over 1680} \gamma_{\rho_1\dots\rho_5} \lambda \, \bar\psi_\mu
\gamma^{\rho_1\dots\rho_5} \epsilon^* \right)\cr
& + {1\over 32} i \left[\left({9\over 4} \gamma_\mu\gamma^\rho +
3\gamma^\rho \gamma_\mu\right)\epsilon\, \bar \lambda
\gamma_\rho\lambda \right.\cr
&\left.   -\left({1\over 24} \gamma_\mu \gamma^{\rho_1\rho_2\rho_3} +
{1\over 6} \gamma^{\rho_1\rho_2\rho_3} \gamma_\mu\right)
\epsilon\,\bar\lambda \gamma_{\rho_1\rho_2\rho_3}\lambda +
{1\over 960} \gamma_\mu\gamma^{\rho_1\dots\rho_5} \epsilon\, \bar
\lambda \gamma_{\rho_1\dots\rho_5} \lambda    \right] \cr
& + \delta^{(0)}_\Sigma (\psi_\mu),
\cr}}
where the compensating $U(1)$  transformation is given by
\eqn\compenpsi{\delta^{(0)}_\Sigma\psi_\mu
= {1\over 2} i \Sigma = {1\over 4} i\psi_\mu
(\bar\epsilon \lambda^*)  - {1\over 4} i
\psi_\mu (\bar\epsilon^* \lambda) .}

By using \fierz\ and \vartino\ extensively we may manipulate the variation of
$\gamma^\mu \psi_\mu^*$   into the form,
\eqn\vargamps{\delta^{(0)}(\gamma^\mu \psi^*_\mu)_a = -{3\over 4}i
\lambda^*_a (\bar \epsilon \lambda) +  {1\over 1920} i
(\gamma^{\rho_1 \dots \rho_5} \epsilon^*)_a  \left(\bar\lambda
\gamma_{\rho_1\dots \rho_5} \lambda \right) + \dots ,}
where we have only kept the terms bilinear in
$\lambda$, $\lambda^*$. This implies the relation,
\eqn\compsi{(\lambda)^{15}_a \, \delta^{(0)}(\gamma^\mu \psi^*_\mu)_a =   -
15 i \lambda^{16} (\bar\lambda \epsilon^*) + \dots,}
which we use in the body of the text.

\appendix{B}{Determination of the Coefficient $c$}

To determine the coefficient $c$ in $L_1^{(0)}$, we need to
consider how this term mixes with other terms under supersymmetry
transformations.  We shall, in particular, consider the term in the dilatino
transformation \lamtran,
\eqn\varuse{\delta^{(0)} \lambda = i\gamma^\mu \epsilon^*\, P_\mu,}
which transforms $L_1^{(0)}$  into the form $\lambda \lambda^{*2}
P_\mu\epsilon^*$.

There are two terms which mix with $L_1^{(0)}$ under this transformation.
One of these, $L_2^{(0)}$,  arises from the $U(1)$
 connection in
 the kinetic term
$\bar\lambda \gamma^\mu D_\mu \lambda$,
\eqn\conlam{L_2^{(0)} = {3\over 2}\, i\, \det\, e\, \bar
\lambda\gamma^\mu \lambda\, Q_\mu.}
It follows from the transformation of $Q_\mu$ in \qtrans\ that  the relevant
transformation of $L_2^{(0)}$ is,
\eqn\ltwot{\delta^{(0)} L_2^{(0)} = -{3\over 2}\, i\, \det\, e
\, \bar\lambda
\gamma^\mu \lambda\, \bar \epsilon\lambda^*\, P_\mu.}

In addition to $L_2^{(0)}$, there is another term in the IIB action
that can be deduced  from  the gravitino equation of motion (eq. (4.12) of
 \schwarza),
\eqn\acterm{L_3^{(0)} = i\,\det\, e\, \bar \lambda
\gamma^\mu\gamma^\omega \psi_\mu^*
 P_\omega.}
The supersymmetry transformation of the gravitino \vartino\ gives the
variation of $L_3^{(0)}$,
\eqn\lthreet{\eqalign{\delta^{(0)} L_3^{(0)} =&  {1\over 32}\, i\, \det\, e\,
\bar \lambda  \gamma^\mu\gamma^\omega  \left[\left({9\over 4}
\gamma_\mu\gamma^\rho +
3\gamma^\rho \gamma_\mu\right)\epsilon^*\, \bar \lambda^*
\gamma_\rho\lambda^* \right.\cr
&\left.   -\left({1\over 24} \gamma_\mu \gamma^{\rho_1\rho_2\rho_3} +
{1\over 6} \gamma^{\rho_1\rho_2\rho_3} \gamma_\mu\right)
\epsilon^*\,\bar\lambda^* \gamma_{\rho_1\rho_2\rho_3}\lambda^* +
{1\over 960} \gamma_\mu\gamma^{\rho_1\dots\rho_5} \epsilon^*\, \bar
\lambda^* \gamma_{\rho_1\dots\rho_5} \lambda^*    \right] P_\omega
\cr
= &  {1\over 32}i\, \det\, e\, \left[12\bar\lambda
\epsilon^*\, \bar\lambda^* \gamma^\omega \lambda^* -\bar\lambda
\left({1\over 3}
\gamma^\omega \gamma^{\rho_1\rho_2\rho_3} + {1\over 3}
\gamma^{\rho_1\rho_2\rho_3} \gamma^\omega\right) \epsilon^* \,
\bar\lambda^* \gamma_{\rho_1\rho_2\rho_3} \lambda^* \right. \cr
& \left.  - {1\over 120}\bar\lambda \gamma^\omega \gamma^{\rho_1\dots
\rho_5}\epsilon^* \, \bar\lambda^* \gamma_{\rho_1\dots\rho_5}
\lambda^* \right] P_\omega .\cr}}

An important simplification occurs when the variations
$\delta^{(0)}L_2$ and $\delta^{(0)} L_3$  are added together by adding
\ltwot\ and \lthreet.  To see this it is first useful to
 use the fundamental Fierz identity,
\fierz, to write
\eqn\applfierz{\eqalign{ A \equiv
\bar\lambda  \gamma^\rho \lambda \, \bar \lambda
\epsilon^* =  & {8\over 9}
\left[{1\over 16} \bar \lambda\gamma^\mu
\gamma^\omega \epsilon^*\, \bar\lambda\gamma_\mu \lambda +
{1\over 96} \bar\lambda\gamma^\omega \gamma^{\rho_1\rho_2\rho_3}
\epsilon^* \, \bar\lambda\gamma_{\rho_1\rho_2\rho_3} \lambda  \right.\cr
&\qquad  \left.-{1\over 240\cdot 16}  \bar\lambda\gamma^\omega
\gamma^{\rho_1\dots\rho_5} \epsilon^* \,
\bar\lambda\gamma_{\rho_1\dots \rho_5} \lambda \right]P_\omega.\cr}}

The sum $ \delta^{(0)}L_2 + \delta^{(0)} L_3$ contains the terms
$(-3/2 + 3/8)i A= - 9i/8 A$.  Substituting in \ltwot\ and \lthreet\ gives
\eqn\tottwo{\eqalign{\delta^{(0)}L_2 + \delta^{(0)} L_3 =
 -{i\over 32}\det\, e\, \left[2
 \bar\lambda \gamma_\mu  \gamma^\omega \epsilon^*\,
\bar\lambda\gamma^\mu \lambda  + {1\over 3}
\bar\lambda
\gamma^{\rho_1\rho_2\rho_3} \gamma^\omega \epsilon^* \,
\bar\lambda^* \gamma_{\rho_1\rho_2\rho_3} \lambda^*  \right] P_\omega .\cr}}
This sum of the variations  has to
cancel the variation of
the term $L_1^{(0)}$ in \nonterm.   To see this most clearly, it
is useful to  first manipulate $L_1^{(0)}$ using \fierz\ into the form,
\eqn\ident{\eqalign{ L_1^{(0)} =&  -\det\, e {c\over 6} \bar\lambda^* 
\gamma^{\mu\nu\rho}
\lambda\, \bar \lambda \gamma_{\mu\nu\rho} \lambda^* \cr
& = \det\, e
 {4 c\over 3}\left(\bar\lambda\gamma^\mu \lambda\, \bar\lambda^* \gamma_\mu
\lambda^* + {1\over 6} \bar\lambda^* \gamma^{\mu\nu\rho} \lambda^* \,
\bar\lambda^* \gamma_{\mu\nu\rho} \lambda^* \right)+ \cdots .\cr}}
Therefore the supersymmetry variation of $L_1^{(0)}$ may be expressed as,
\eqn\lonet{ \delta^{(0)}L_1^{(0)} = \det\, e\,
 {8 c\over 3}i \left(\bar\lambda^*\gamma^\mu \lambda^*\,
 \bar\lambda \gamma_\mu  \gamma^\omega \epsilon^*  +
 {1\over 6}\bar\lambda^* \gamma^{\mu\nu\rho} \lambda^* \,
 \bar\lambda \gamma_{\mu\nu\rho} \gamma^\omega\epsilon^*
\right)P_\omega + \cdots  ,}
which can be compared directly with \tottwo.
In order for the sum of \tottwo\ and \lonet\ to vanish
the coefficient $c$ must  have the value,
\eqn\cresult{c=-{3\over 128}.}

\footatend\vfill\supereject\immediate\closeout\rfile\writestoppt
\baselineskip=14pt\centerline{{\bf References}}\bigskip{\frenchspacing%
\parindent=20pt\escapechar=` \input refs.tmp\vfill\eject}\nonfrenchspacing
\bye